\documentclass[12pt,preprint]{aastex}

\def\Msun{\hbox{$\rm\thinspace M_{\odot}$}}
\def\farcs{\hbox{$.\!\!^{\prime\prime}$}}
\def\arcsec{\hbox{$^{\prime\prime}$}}
\newcommand\tabspace{\noalign{\vspace*{0.7mm}}} 
\def\errtwo#1#2#3{$#1^{+#2}_{-#3}$}

\newcommand\aproxgt{\mathrel{%
      \rlap{\raise 0.511ex \hbox{$>$}}{\lower 0.511ex \hbox{$\sim$}}}}
\newcommand\aproxlt{\mathrel{%
      \rlap{\raise 0.511ex \hbox{$<$}}{\lower 0.511ex \hbox{$\sim$}}}}

\shorttitle{A simultaneous broadband campaign on M81*}
\shortauthors{Markoff et al.}

\begin{document}

\title{Results from an extensive simultaneous broadband campaign on
  the underluminous active nucleus M81*: further evidence for
  mass-scaling accretion in black holes }

\author{Sera Markoff}
\affil{Sterrenkundig Instituut ``Anton Pannekoek'', University of
  Amsterdam, 1098 SJ Amsterdam, the Netherlands}
\email{sera@science.uva.nl}

\and

\author{Michael Nowak, Andrew Young\altaffilmark{1}, Herman L. Marshall, Claude
  R. Canizares}
\affil{Massachusetts Institute of Technology, Kavli Institute for
  Astrophysics and Space Research, Cambridge, MA 02139}
\email{mnowak,ayoung@space.mit.edu}

\and

\author{Alison Peck\altaffilmark{2}} 
\affil{Harvard-Smithsonian Center for Astrophysics, Submillimeter Array Project,
Hilo, HI 96720 and Joint ALMA Office, Santiago 7550108, Chile}
\email{apeck@alma.cl}
\and

\author{Melanie Krips and Glen Petitpas}
\affil{Harvard-Smithsonian Center for Astrophysics, Submillimeter
  Array Project, Hilo, HI
  96720}\email{mkrips,gpetitpas@cfa.harvard.edu}

\and

\author{Rainer Sch\"odel}
\affil{Instituto de Astrof\'isica de Andaluc\'ia (CSIC),18008 Granada, Spain}
\email{rainer@iaa.es}

\and

\author{Geoffrey C. Bower}
\affil{Department of Astronomy, University of California,  
Berkeley, CA 94720}
\email{gbower@astro.berkeley.edu}

\and

\author{Poonam Chandra\altaffilmark{3}} 
\affil{Astronomy Department,
University of Virginia, Charlottesville VA 22903
and National Radio Astronomy Observatory,
Charlottesville, VA 22903} 
\email{ pchandra@virginia.edu}

\and
 
\author{Alak Ray}
\affil{Tata Institute of Fundamental Research, Mumbai 400 005, India}
\email{akr@tifr.res.in}

\and 

\author{Michael Muno\altaffilmark{4}, Sarah Gallagher\altaffilmark{5}
  and Seth Hornstein \altaffilmark{6}}
\affil{University of California, Los Angeles, Physics \& Astronomy
Building, Box 951547, Los Angeles, CA 90095}
\email{mmuno,sgall,seth@astro.ucla.edu}

\and 

\author{Chi C. Cheung\altaffilmark{7}}
\affil{NASA Goddard Space Flight Center, Greenbelt, MD, 20771}
\email{chi.c.cheung@nasa.gov}

\altaffiltext{1}{Current affiliation: University of Bristol, Bristol, UK}
\altaffiltext{2}{Current affiliation: NRAO, Joint ALMA Office, Chile}
\altaffiltext{3}{NRAO Jansky Fellow}
\altaffiltext{4}{Hubble Fellow, currently at California Institute of Technology}
\altaffiltext{5}{Spitzer Fellow, currently at University of Western
  Ontario, London, ON, Canada}
\altaffiltext{6}{Current affiliation: Center for Astrophysics \& Space
  Astronomy, University of Colorado at Boulder}
\altaffiltext{7}{NASA Postdoctoral Fellow}

\begin{abstract}

We present the results of a broadband simultaneous campaign on the
nearby low-luminosity active galactic nucleus M81*.  From February
through August 2005, we observed M81* five times using the
\textit{Chandra X-ray Observatory} with the High-Energy Transmission
Grating Spectrometer, complemented by ground-based observations with
the Giant Meterwave Radio Telescope, the Very Large Array and Very
Large Baseline Array, the Plateau de Bure Interferometer at IRAM, the
Submillimeter Array and Lick Observatory. We discuss how the resulting
spectra vary over short and longer timescales compared to previous
results, especially in the X-rays where this is the first ever
longer-term campaign at spatial resolution high enough to nearly
isolate the nucleus (17pc).  We compare the spectrum to our Galactic
center weakly active nucleus Sgr A*, which has undergone similar
campaigns, as well as to weakly accreting X-ray binaries in the
context of outflow-dominated models.  In agreement with recent results
suggesting that the physics of weakly-accreting black holes scales
predictably with mass, we find that the exact same model which
successfully describes hard state X-ray binaries applies to M81*, with
very similar physical parameters.  
  
\end{abstract}

\keywords{Galaxies: active---galaxies: individual: Messier Number:
  M81 --- radiation mechanisms: non-thermal---black hole
  physics---accretion, accretion disks}

\section{Introduction}

Black holes are a common feature in galaxies, spanning a huge range in
mass, from the stellar-sized remnants scattered throughout the galaxy
volume to the supermassive black hole often thought to lurk in the
galaxy center.  Black holes are known to operate over at least ten
orders of magnitude in luminosity, and thus experience accretion rates
that range from super-Eddington (defined with respect to an associated
Eddington luminosity via $\dot M_{\rm Edd} \equiv L_{\rm Edd}/c^2$,
where $L_{\rm Edd}\equiv {4\pi c G M m}/{\sigma_{\tau}}$) to extreme
sub-Eddington.  Clearly the accretion rate is the overall dominating
factor determining the energy output; however, the accretion flow
behavior of at least stellar mass black holes changes rather
drastically between low to high accretion rates.  Variations include
the cyclic appearance and apparent subsequent quenching of jet
outflows, alterations in accretion disk characteristics, and changes
in the overall radiative efficiency.  For a description of the
  most recent observations and their implications for theoretical
  models of accretion flows, see the reviews by
  \cite{RemillardMcClintock2006,DoneGierlinskiKubota2007}, and
  references therein.

Above a few percent of the Eddington luminosity, for instance,
accretion around black holes seems to be well-characterized by a
dominant, thermally emitting ``standard thin disk''
\citep{ShakuraSunyaev1973}. Somewhere below this threshold there
appears to be a transition to a radiatively inefficient state with
some form of advective accretion and/or outflow \citep[see,
e.g.,][]{MeyerMeyerHofmeister1994,NarayanYi1994,BlandfordBegelman1999,QuataertGruzinov1999}.
Although this transition has been predicted theoretically, the
physical details and configurations of weakly accreting flows are
still under significant debate \citep[see, e.g.][]{Rykoffetal2007}.
In particular there are open questions regarding fundamental plasma
characteristics such as the coupling-- and therefore respective
temperatures-- of the ions and electrons, viscosity and the role of
magnetic fields, the accretion flow geometry, and the relationship of
the accretion flow to outflows and jet production.

For stellar mass black holes, we are learning a great deal by
observing the transitions in real time between accretion states.  
Unfortunately, it is not possible to track such changes directly in
supermassive black holes (SMBHs) as the relevant dynamical times scale
approximately with the mass. Instead, ensemble comparisons can be made
with large samples of accreting SMBHs that range from near-Eddington
down to the intrinsically weakest active galactic nuclei (AGN), called
low-luminosity AGN (LLAGN; \citealt{Heckman1980,Ho1999}).  These
sources are difficult to observe unless nearby, however, because of
their intrinsically weak emission.

Given its special role as the weakest observable active nucleus, Sgr
A*, our Galactic center SMBH, has become the poster-child for a
multitude of theoretical and observational studies.  Several extensive
multiwavelength campaigns
\citep[e.g.][]{Baganoffetal2003,Eckartetal2004,Anetal2005,Yusef-Zadehetal2006}
have well established the simultaneous broadband spectrum of Sgr A*,
which provides a tight constraint on physical models.  Over the course
of these campaigns, however, some very unusual flaring behavior has
been discovered in Sgr A*'s X-ray emission.  No other black hole has
shown similar flaring; however, currently we are unable to detect any
other Sgr A*-like objects for comparison.  Furthermore, the presence
of jets has not been definitively confirmed or ruled out for Sgr A*,
complicating our ability to use it as a test source for
accretion/outflow relationships at low accretion rates \citep[but
see][]{MarkoffBowerFalcke2007}.  The distinct lack of any thin
accretion disk component in its spectrum also makes Sgr A* unique
compared to other LLAGN.

What seems to be required is a ``bridge'' source, to help span the gap
between Sgr A* and other LLAGN in terms of accretion rate, while
sharing as many other qualities as possible (mass, spectral features,
galaxy type, etc.).  Comparing Sgr A* to such a source would help us
determine what processes may be different or absent at the lowest
accretion rates.  A comparison of this type, however, would require
simultaneous broadband data for the bridge source, of the quality now
only associated with the Galactic center campaigns.

The nucleus of the nearby galaxy M81 is an ideal candidate for such an
extensive multiwavelength campaign on a LLAGN, for the following
reasons.  Although it is one of the intrinsically weakest LLAGN known,
it is among the brightest because it is the nearest galaxy besides
Centaurus A with a central AGN. Furthermore it is the nearest
point-like LLAGN, similarly inside a spiral galaxy, for which reliable
measurements of the black hole mass are available.  Spectroscopy with
the \textit{Hubble Space Telescope (HST)} suggests a central mass of
$7\times10^7$\Msun\ \citep{Devereuxetal2003}, and M81's distance is
only 3.6 Mpc \citep{Freedmanetal1994}.  The nucleus, M81* (following
the convention based on Sgr A*), is associated with a compact radio
core and exhibits both low-ionization emission line region (LINER;
\citealt{Heckman1980,Ho1999}) and Seyfert 1 characteristics.  In terms
of radiative power, probable accretion rate, and the length of the jet
emanating from the core, M81* lies in the intermediate range between
radio loud AGN and Sgr A*.

The X-ray properties of M81* are very typical for the LLAGN class
\citep[e.g.][]{Ho1999}.  Its nonthermal X-ray luminosity is around a
few $10^{-5} L_{\rm Edd}$, though its proximity means it is still
bright.  Similarly its spectral energy distribution (SED) displays no
``big blue bump'' yet does show evidence for double-peaked optical
line emission \citep{Boweretal1996}, suggesting the presence of a
weak accretion disk.  {\em HST} observations with STIS indicate that
the disk is close to face-on with an inclination of 14$^\circ$
\citep{Devereuxetal2003}.

As will be described in more detail below, M81* exhibits significant
variability across its SED on both short (daily) and long
(monthly/yearly) time scales.  M81* has a low absorption column
\citep{Pageetal2003}, which allows UV and soft X-ray detection.
Perhaps most importantly for the goals of this campaign, M81* shares several
important characteristics with Sgr A*, specifically its radio slope and
polarization, that make it the ideal comparison source.  
As studies of our own Galactic center have shown, detailed,
multi-wavelength observations of single objects such as M81* are
indispensable tools for understanding black hole accretion.

In this paper we will present the results of a simultaneous,
broadband, multiwavelength campaign on M81*, and discuss how it
compares with Sgr A*, as well as its weakly accreting black hole
counterparts in X-ray binaries (XRBs).  The {\em Chandra} observations
specifically resulted in the first gratings-resolution X-ray spectrum
of an isolated LLAGN nucleus, which is the focus of a companion paper
\citep{Youngetal2007}.  The millimeter observations carried out with
the Plateau de Bure Interferometer(PdBI) are also presented in more
detail elsewhere \citep{Schoedeletal2007}. In Section~\ref{sec:oldobs}
we summarize the results of previous observations of M81* and in
Section~\ref{sec:obs} we describe our new observations and analysis.
We present the resulting broadband spectra in Section~\ref{sec:newobs}
and their interpretation in the specific context of a jet-dominated
model in Sections~\ref{sec:model} \& \ref{sec:results}.
Section~\ref{sec:discuss} contains our discussion and conclusions.

\section{Previous Observations of M81*}\label{sec:oldobs}

The M81* nucleus has been observed extensively in many wavebands
(partly due to its proximity to supernova SN 1993J, which undergoes
regular monitoring).  In this section we will briefly summarize the
previously known broadband characteristics of M81* and discuss how
they have provided the motivation for more detailed and higher
resolution observations.

\subsection{Radio/millimeter Observations} \label{subsec:radio}

M81* exhibits the signature flat/inverted radio spectrum associated
with the compact cores of AGN.  Such a spectrum is well-explained by a
collimated, steady jet which radiates via self-absorbed synchrotron
emission along its length \citep[see,
e.g.][]{BlandfordKoenigl1979,Falcke1996}.  Based on observations from
1.4--22.5 GHz using the VLA, several groups have observed an inverted
spectrum ($\alpha\sim 0.0$--0.3, $F_\nu\propto\nu^{\alpha}$)
with a flux in the range of $\approx80-300$ mJy
\citep{Hoetal1999,BietenholzBartelRupen2000,Brunthaleretal2001,BrunthalerBowerFalcke2006}.
Based on four years of observations with the VLA, \citet{Hoetal1999}
note that M81* undergoes frequent radio outbursts, with the underlying
flux around  $100$\,mJy and higher fluxes during flares.
The larger flare events occur on time scales of months, and seem to
roughly correspond with predictions of simple adiabatic expansion
models \citep[e.g.][]{vanderLaan1966} in which the variability moves
towards lower frequencies as the flare amplitude decreases.  Ho et
al. also claim they detect intraday variability at a level of
10-60\% amplitude changes.  If the longer term radio flares are indeed
expanding ejecta moving out along otherwise steady jet structures,
there should be even higher amplitude variations in the millimeter
regime.  Consistent with this view, \cite{Sakamotoetal2001} observed
the 3 mm flux to double within a single day.

A one-sided jet system has been resolved in M81*.
\cite{BietenholzBartelRupen2000} identified a stationary radio core
with a very small (700 AU at 22 GHz), precessing (over 20$^\circ$)
jet, the structure of which varies on relatively short timescales.
Interestingly, this group found no significant intraday variations
over a similar time frame as the Ho et al. study.  Investigating this
question of variability was one of our campaign's goals for the lower
frequencies.

M81*'s radio luminosity is about four orders of magnitude brighter
than Sgr A*'s, but its shape and polarization are quite similar.  A
key, and somewhat unusual, trait that these two sources share is the
dominance of circular polarization up to $\aproxgt 22$\,GHz in M81*,
and $\sim 112$\,GHz in Sgr A*
\citep{Brunthaleretal2001,Boweretal2002,BrunthalerBowerFalcke2006}. Sgr
A* becomes increasingly linearly polarized towards its peak flux in
the submm \citep{Boweretal2003,Boweretal2005}; however, the
characteristics of M81* in the submm have not yet been determined.
The interpretation of Sgr A*'s circular polarization is Faraday
depolarization by the surrounding accretion flow.  If the same physics
is active in M81*, we would expect to detect linear polarization in
the submm range as well.

\subsection{Infrared through Ultraviolet Observations} \label{subsec:uv}

Typical of LLAGN, M81* lacks a bright optically thick ``standard thin
disk'' component in the optical range, although double-peaked optical
lines do suggest the presence of a weak disk \citep{Boweretal1996}.
The low column ($N_h\sim5\times10^{20}$ cm$^{-2}$;
\citealt{Pageetal2003}) allows us to detect UV and soft X-rays from
the nucleus, which suggests a nearly face-on disk.
\cite{Maozetal2005} used the {\em HST ACS} to discover variable (by
tens of \%) UV emission, consistent with the results of
\cite{HoFilippenkoSargent1996}, who detected a weak and very steep
($\alpha\approx-2$) UV continuum.  Instruments with less spatial
resolution such as {\em Spitzer}
\citep{Willneretal2004,Murphyetal2006}, MIRLIN \citep{Grossanetal2001}
and ISOPHOT-S \citep{Satyapaletal2005} are all consistent with a steep
($\alpha\le-1.7$) non-stellar spectrum, similar to what is observed in
the IR/UV in the LLAGN NGC~4258 \citep{Charyetal2000}.  This is of
interest because it suggests that in LLAGN, the UV is likely
nonthermal emission, while the optical, and perhaps IR, contain the
only potential  signatures of a radiatively efficient,  thin accretion disk.

\subsection{X-ray Observations} \label{subsec:xray}

M81* has a persistent nonthermal power-law flux in the X-rays, with
variations of factors of 3 or more over yearly time scales.  
{\em ASCA} has detected both long term variations, as well as 20--30\%
intraday variations which suggest that the source size is less than a
few hundred gravitational radii ($r_{\rm g}\equiv GM/c^2$)
\citep{IyamotoMakishima2001}.  {\em ROSAT} also confirmed long term
X-ray variability, with a factor of $\approx 2.5$ amplitude
\citep{ImmlerWang2001}.  A summary of these and more recent
variability trends can be found in \cite{LaParolaetal2004}.  M81*'s
X-ray luminosity appears to vary between ($2$--$6)\times10^{41}$ ergs
s$^{-1}$, which corresponds to $\sim (2$--$6) \times 10^{-5} L_{\rm
Edd}$.  It was not until observations with {\em BeppoSAX} that more
detailed statements could be made about the nuclear X-ray emission
properties.  \cite{Pellegrinietal2000} observed both the short
intraday as well as the long term variability over the 0.1--100\,keV
band.  However, due to the poor angular resolution, they could only
place a lower limit of $\aproxgt80\%$ of the continuum originating in the
nucleus.

The {\em BeppoSAX} observations yielded several important new results
which contributed to our interest in M81* as a potential {\em Chandra HETGS}
target.  First, the data were consistent with no reflection component
or blue bump, which would be unusual for a Seyfert 1 -- a class of
objects with which M81* otherwise shares some qualities. The {\em
BeppoSAX} results make it less likely that M81* is a simple extension
of the Seyfert 1 class to low luminosity.  Although {\em BeppoSAX} did
not detect reflection, it did detect emission and absorption features
of highly ionized iron; however, these features were seemingly not
correlated with the continuum luminosity.  Second, there were problems
reconciling the ionization with the low inferred accretion rate.
\cite{Pellegrinietal2000} suggested that instead of ionization, there
may instead be transmission through a highly photoionized medium close
to the nucleus, such as a warm absorber.  Third, while the observed
X-ray powerlaw with $\Gamma\approx1.8$--1.9 was typical of bright
Seyferts, there was no direct evidence for a thin accretion disk.  A
lack of a strong disk component is consistent with accretion being
dominated by a radiatively inefficient accretion flow (RIAF); however,
the question then arises whether such flows can account for the
strong Seyfert-like powerlaw over the entire 0.1-100\,keV range at the
low inferred accretion rates of M81*.  {\em XMM-Newton} has confirmed
these findings, at least in the 0.3-8\,keV band, and also detected
redshifted Fe K$ \alpha$ as well as He- and H-like ionized iron
\citep{Pageetal2003}.

Because {\em Chandra} is the only X-ray mission with the spatial
resolution to almost isolate the nucleus of M81* (to within $\sim 17$
pc of the black hole), determining the nature of the line emission was
one of our primary goals.  As described in \cite{Youngetal2007}, we
indeed detect not just iron but many other low-metallicity species, as
well as velocity broadening of some of these lines.  The broadened
line components are consistent with arising from regions close to the
black hole, i.e., $\aproxlt 10^{5}~GM/c^2$.  Detected line features
include those associated with fluorescence from cold material (Ar
K$\alpha$, Si K$\alpha$, and Fe K$\alpha$), emission lines from a hot
plasma (Ne {\sc x}, Mg {\sc xii}, Si {\sc xiii}), and absorption lines
(18.44\,\AA\ and 20.74\,\AA) that could be consistent with an
outflowing wind.  The plasma emission lines, specifically the Si {\sc
xiii} triplet line strength ratios, are consistent with a
collisionally ionized plasma (although other models can not be ruled
out).  The focus of the \cite{Youngetal2007} work is on the X-ray
spectra of M81*, specifically the aforementioned line features in
relation to the X-ray continuum.  Young et al. show that the X-ray
spectra are consistent with the expectations of a somewhat simplified
RIAF model.  We did not consider the simultaneous radio and submm data
in the context of those models.  In this work we now include a
description of the lower frequency (radio through sub-millimeter)
observations, and consider the entire broadband spectrum in the context
of an outflow-dominated model.

\section{Observations and Data Reduction}\label{sec:obs}

The difficulty in observing a source as faint as M81* with the {\em
Chandra HETGS} lies in the required long integration times (300\,ksec)
to adequately resolve the narrow ($\aproxlt 2500~{\rm km~s^{-1}}$ Full
Width Half Maximum) line features.  Additionally, aside from the Fe
features found with {\em BeppoSAX} and {\em XMM} observations (which,
due to the relatively poorer spatial resolution of these instruments,
could have arisen from well outside the nucleus), the existence of
line features from the innermost regions of LLAGN was uncertain. As
such, at the time of our M81* observations, no gratings observation of
an LLAGN had been accepted in the {\em Chandra} Guest Observer
program.  Instead, we obtained a series of {\em HETGS} observations
via the Guaranteed Time Observation program (PI: C. Canizares). In
order to further constrain models and better understand the
variability trends of M81*, we supplemented the {\em Chandra} program
by proposing for simultaneous coverage with five ground-based
instruments that span the lower frequencies: the Giant Meterwave Radio
Telescope (GMRT), the Very Large Array/Very Large Baseline Array
(VLA/VLBA), the Plateau de Bure Interferometer (PdBI) at IRAM, the
Sub-millimeter Array (SMA), and Lick Observatory.

\begin{figure*}
\epsscale{1}
\plotone{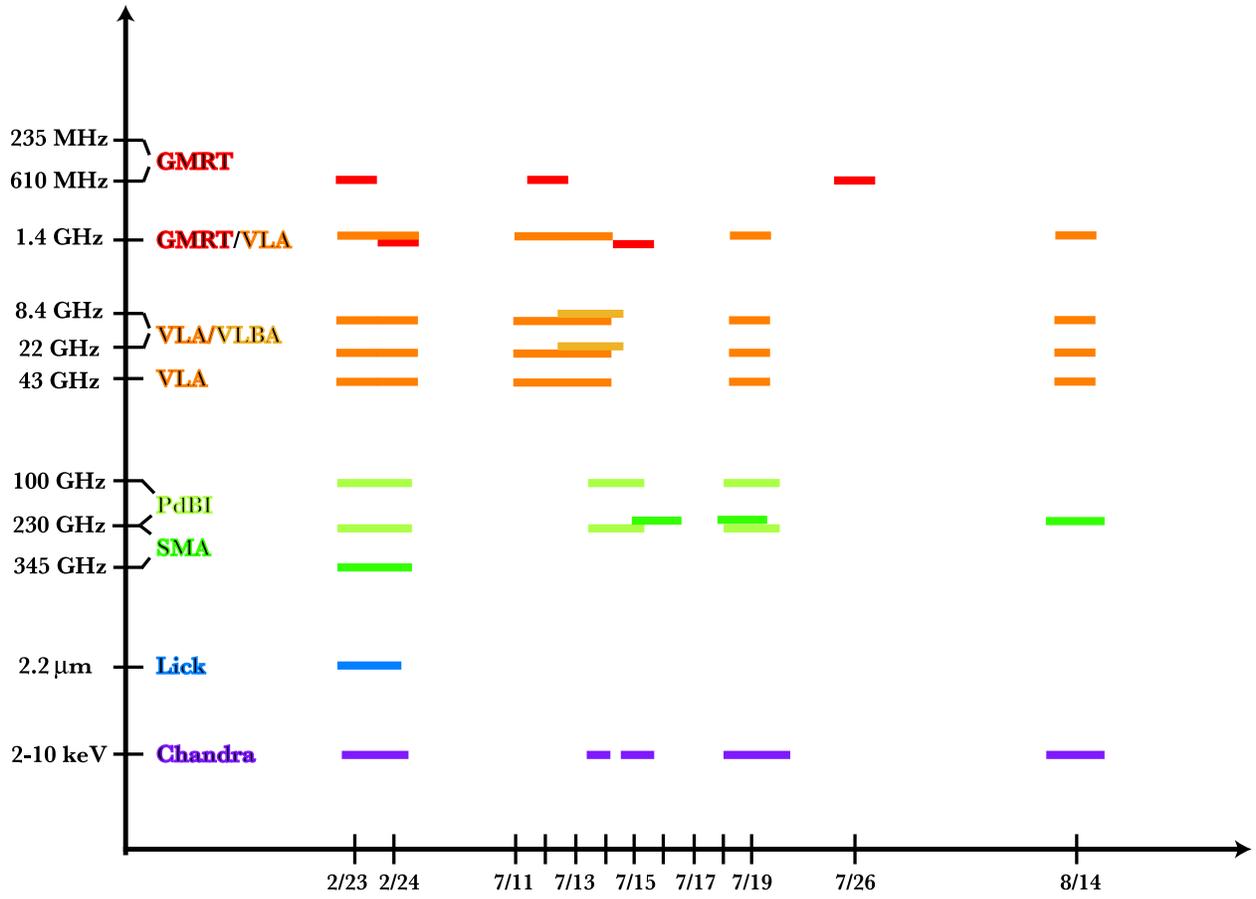}
\caption{Schematic overview of the entire campaign during 2005.\label{campoview}}
\end{figure*}

\begin{figure*}
\epsscale{1}
\plotone{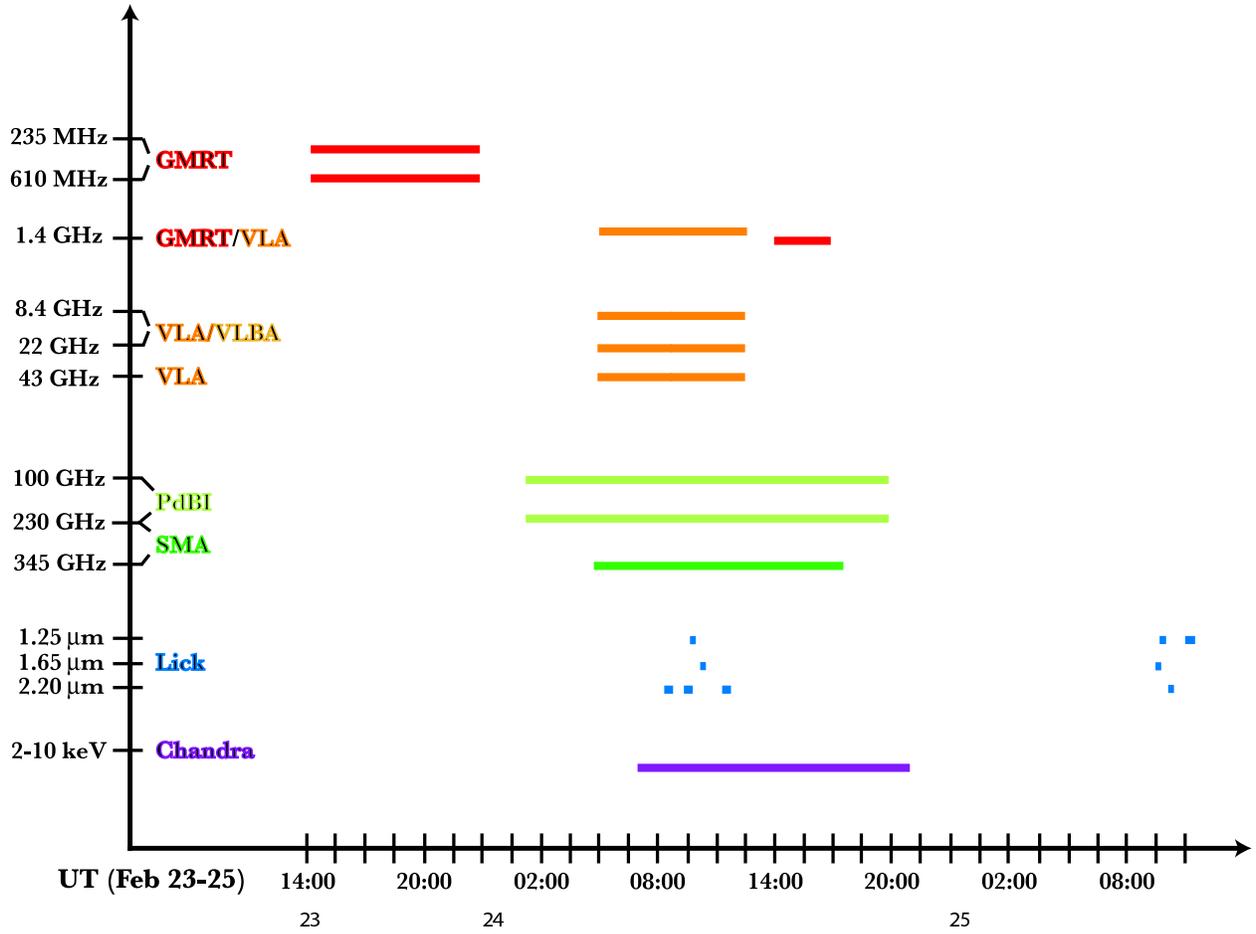}
\caption{Schematic overview of the 2005 February 23--25 period.
  Exact times are shown in Table~\ref{febtimes}.\label{febview}}
\end{figure*}

\begin{figure*}
\epsscale{1}
\plotone{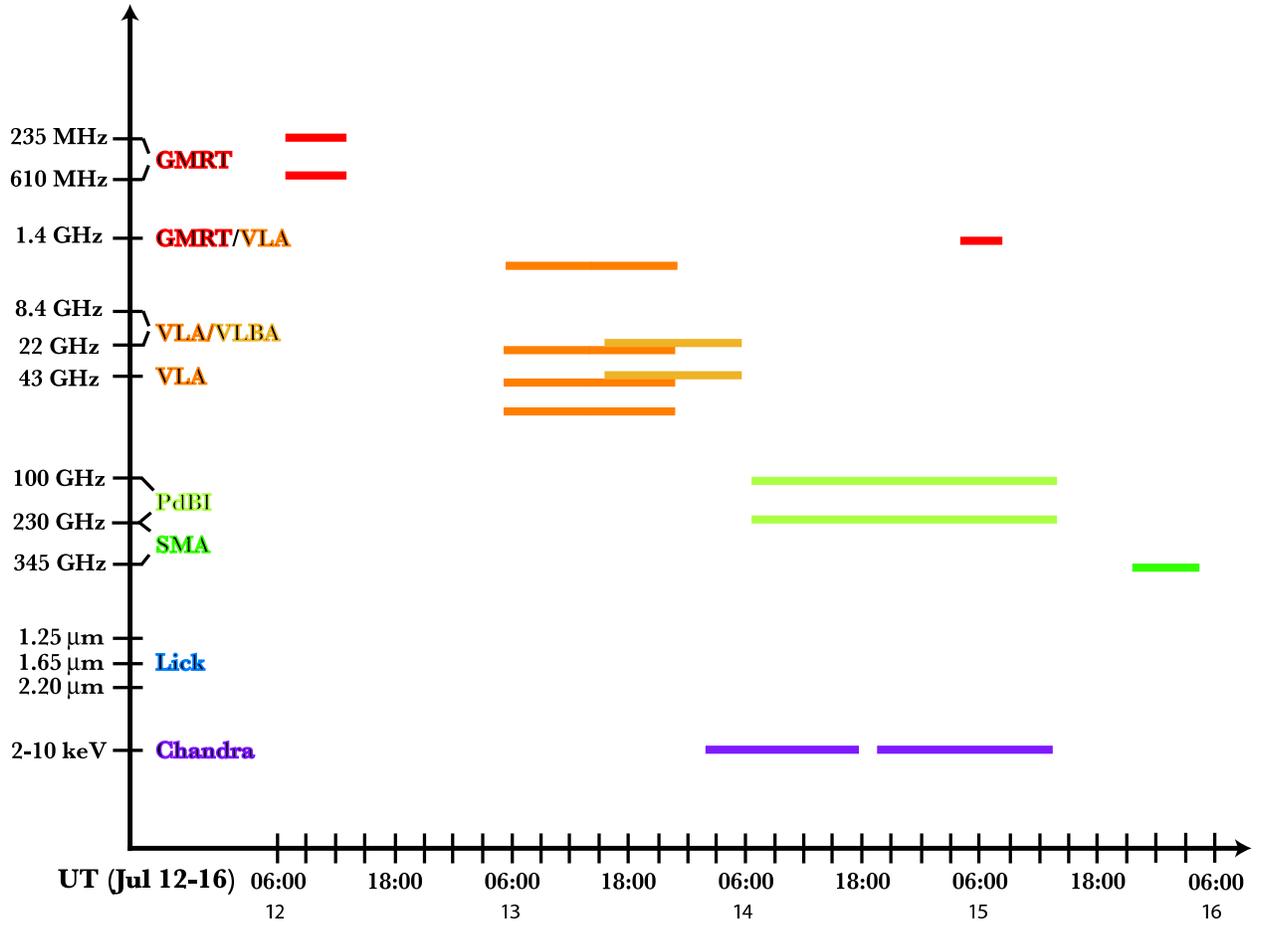}
\caption{Schematic overview of the 2005 July 12-16 period.  Exact
  times are shown in Table~\ref{jultimes}. 
    \label{julview}}
\end{figure*}

Figure~\ref{campoview} gives an overview of the total campaign.  For
the two periods of greatest multi-wavelength overlap,
Figures~\ref{febview} \& \ref{julview} give a closeup view of the
coverage, and Tables ~\ref{febtimes}-- \ref{augtimes} give the exact
times in UT.

In the following subsections we provide details of the individual
instrument observations and data reduction.

\subsection{Low-frequency Radio Waves:  GMRT} \label{subsec:gmrt}

The Giant Meterwave Radio Telescope (GMRT) is an aperture synthesis
radio telescope \citep{Swarupetal1991} situated 80 km north of Pune
in Western India at latitude $19^o \,06'$ and longitude $74^o\,03'$ E. The
telescope operates at 233, 327, 610 and 1420 MHz bands, and consists
of 30 fixed position, fully steerable paraboloid dishes of diameter 45
meters. Fourteen out of these 30 dishes are located within about a
square kilometer of each other, the remaining 16 antennae form a
``Y''-shaped array with northwest, northeast and southern arms spread
over an area of 25 kilometers in diameter. The baselines in the
central one square kilometer area are useful to map the extended
emission of the source, whereas the wider baselines in the ``Y''
provide high angular resolution.	

We observed M81* with the GMRT in 1420, 610 and 235 MHz bands on
several occasions during the campaign. The total time spent on M81* in
the 235/610 bands was 5--8 hours, and 3--5 hours in the 1420 MHz band.
The bandwidth at 1420 and 610 MHz was 16 MHz, divided into a total of
128 frequency channels, i.e., the default for the correlator.  For the
243 MHz wave band the bandwidth was 6 MHz.

Calibrator sources were used to remove the effects of
instrumental variations in the measurements. 3C48, 3C286 and
3C147 were used as flux calibrators.  1035+564 was used as a phase
calibrator in the 1420 MHz observations, whereas 0834+555 was used in
the 610 and 235 MHz observations. The flux and phase calibrators were used
for bandpass calibration as well. Flux calibrators were observed once
or twice for 20--30 minutes during each observing session. Phase
calibrators were observed for 5--6 minutes after every 25 minutes of
observations.

We used the Astronomical Image Processing System (AIPS) developed by
NRAO for the data analysis, including the standard GMRT data reduction
\cite[see][for details]{ChandraRayBhatnagar2004}.  Standard flagging
routines of AIPS were used to remove the bad antennas and corrupted
data. About 25--30 antennae could be used in the radio interferometric
setup at different observing epochs. Data were then calibrated and
images and fields were formed by Fourier inversion and CLEANing using
AIPS task IMAGR. We took into account the bandwidth smearing effects
and wide field imaging, even though in this case M81* is not spatially
resolved, as a byproduct of analysis to derive the correct flux
density of the nearby SN 1993J.  Bandwidth effects were negligible for
1420 and 610 MHz bands and we averaged 100 central channels.  For the
235 MHz observations, we divided the central 55--60 good channels and
divided them into 4 sub-bands and stacked them together while
imaging. To take care of the wide field imaging, we divided the whole
field in 3 subfields for 1420 MHz and 610 MHz observations, and into
18 subfields for 235 MHz observations. A few rounds of
self-calibrations were also performed in all the datasets to remove
the phase variations.  AIPS task FLATN was used to combine all the sub
fields into one single image. Table~\ref{m81_obs} gives details of the
observations.  The typical resolution of $3"$ at 1390 MHz at a
distance of 3.6 Mpc for M81 corresponds to about 54 pc.
Table~\ref{sn1993} shows a comparison of calibration observations of
SN 1993J for both the GMRT and the VLA, indicating consistent flux
levels.

\subsection{Centimeter Radio:\\ VLA and VLBA+Effelsberg} \label{subsec:vla}

The Very Large Array observed M81* on 2005 14 February, 13 July, 19 July,
and 14 August.  Observations were obtained at 1.4, 8.4, 22, and
43 GHz on each of the days in continuum mode.  Data were obtained in
fast-switching mode between M81* and the compact calibrators J1048+717
and J1056+701.  We performed calibration of amplitude and phase
variations on short time-scales using J1048+717 and transferred
solutions to M81* and J1056+701.  Results for J1056+701 are,
therefore, a check on variability of M81*.  The amplitude scale was
set by observations of 3C 286.  Weather on 14 February and 13 July was
poor making results at 22 and 43 GHz inaccurate.  We determined
average flux densities for each day as well as measuring flux density
on short time scales.  SN 1993J was in the field of view at 1.4 and
8.4 GHz.  Its flux density was constant between the epochs.

\begin{figure*}
\epsscale{1} 
\plotone{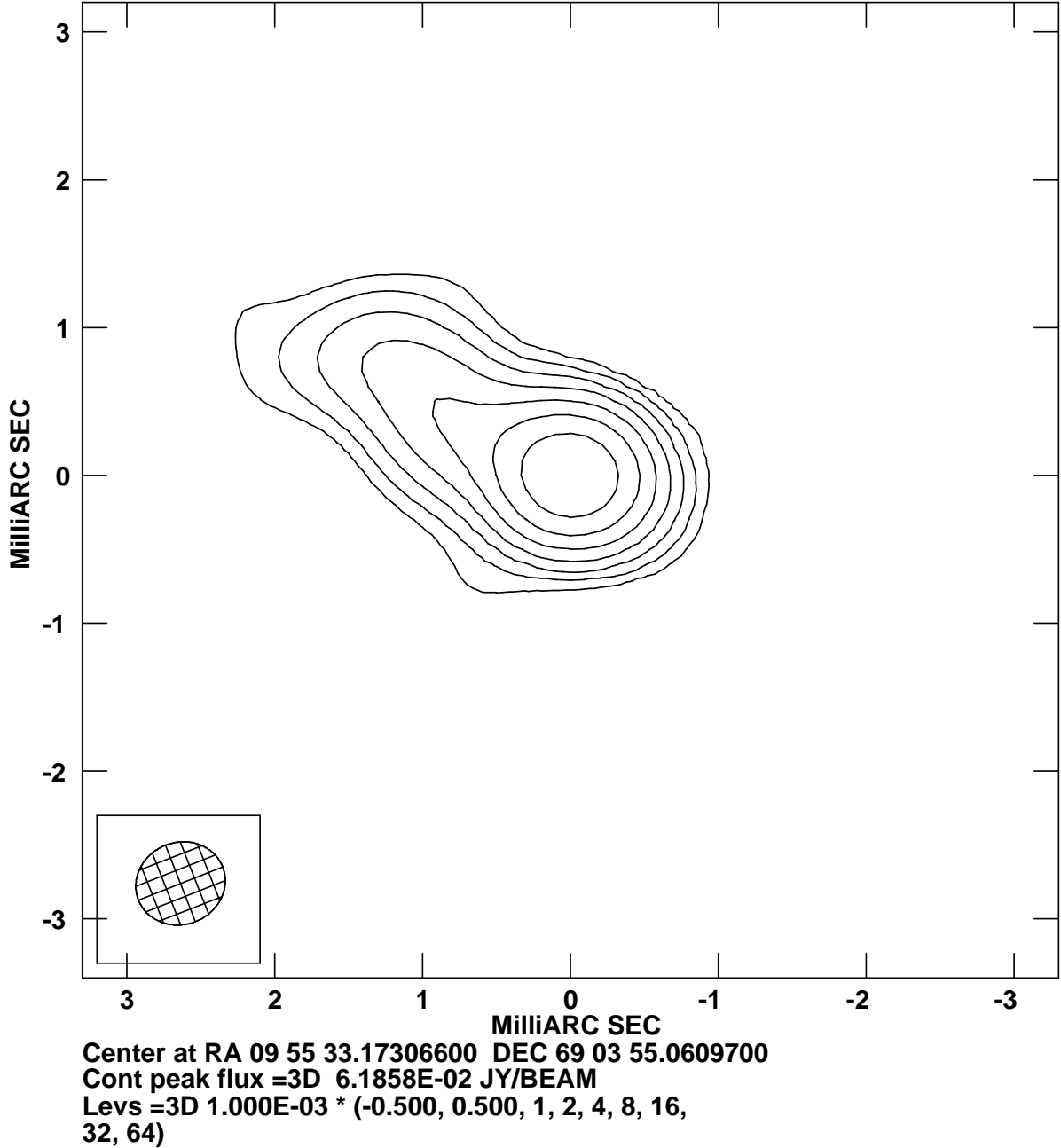}
\caption{Resolved core-jet structure from the 8.4 GHz VLBA+Effelsberg
observations.The synthesized beam of 0.6 mas is shown in the lower
left hand corner.  Contours are 0.5, 1, 2, 4, 8, 16, 32, and 64
mJy.\label{vlbafig}}
\end{figure*}

The Very Long Baseline Array and the Effelsberg 100m observed M81* on
2005 13 July.  Observations were made at 8.4 GHz with a sampling rate
of 128 Mb/s, while attempts at 22 GHz failed due to poor weather.
Standard self-calibration reduction techniques were performed and M81*
was imaged with a resolution of 0.6 mas.  M81* is clearly resolved
into a compact and extended component (Fig.~\ref{vlbafig}).  The peak
flux density in the compact component is about 60 mJy.  In the
extended component, the peak flux density is about 10 mJy, but the
total flux density in the whole region is $\approx2$--3 times that
value.  We fit the image with two elliptical Gaussians: the best fit
compact component has major axis of 0.65\,mas, a minor axis of 0.57\,mas
and a position angle of 81$^\circ$, and the best fit extended
component has major axis 1.5 mas, minor axis 0.66\,mas and position
angle 55$^\circ$.  The separation between the two components is
approximately 1 mas, and the total extension is about $10^4$ AU
across.  These results are similar to what was observed by
\cite{BietenholzBartelRupen2000}.

Our results are summarized in Table~\ref{vla_obs}.  We obtained limits
on the linear polarization on 2005 13 July of 0.1\% at all
frequencies, but did not obtain accurate limits on the circular
polarization of the source.

\subsection{Millimeter Radio: PdBI} \label{subsec:pdbi}

The continuum radiation from M81* at wavelengths of $\sim$3\,mm and
$\sim$1\,mm was observed with the IRAM Plateau de Bure Interferometer
(PdBI) on 2005 24 February, 14-15 July, and 19-20 July. The
observation frequencies differ slightly between the individual epochs
because they were fine-tuned in order to optimize phase stability
depending on the weather conditions. A detailed description of the
data, their reduction and calibration can be found in
\cite{Schoedeletal2007}.  The systematic absolute uncertainty of the
flux calibration is 10-15\% for the 3mm data, and 15-30\% for the 1mm
data (see Table~2 in Sch\"odel et al.).  Flux measurements were
extracted from individual scans of 20\,min duration and detailed light
curves were obtained that show significant variability of M81* during
the observations.  If we add the systematic uncertainties with
  the statistical uncertainties in quadrature,the resulting overall
  uncertainty is around 20\% (3mm) and 30\% (1mm).  Because the
  detected flux variations are likely real,and fully consistent with
  what is observed in other wavebands, including the systematics in
  this manner would seem to overestimate the actual uncertainties.
  For this reason we believe that taking the standard deviation of the
  flux is more representative of the total uncertainty in the context
  of this broadband system.  For this reason we use
the average fluxes and their standard deviations as obtained from the
light curves presented in Sch\"odel et al. (see their Table~3), which
we list in Table~\ref{tab:pdbi}.

The PdBI observed significant variability between the individual
observing epochs, as well as intraday variability. The 3mm and 1 mm
light curves from 2005 24 February --- actually the best of the data
obtained with the PdBI during the campaign --- show a flux decrease
with a significance of $>5\sigma$ over 5 hours that occurred at both
wavelengths (see \citealt{Schoedeletal2007} for a detailed
discussion). The lightcurves are presented in combination with those from the
Submillimeter Array below in Figs.~\ref{febpdbsma} \& \ref{1mmpdbsma}.

\subsection{Submillimeter:  SMA} \label{subsec:sma}

M81* was observed at the Submillimeter Array\footnote{The
Submillimeter Array is a joint project between the Smithsonian
Astrophysical Observatory and the Academia Sinica Institute of
Astronomy and Astrophysics, and is funded by the Smithsonian
Institution and the Academia Sinica.} (SMA; \citealt{HoMoranLo2004}) on
Mauna Kea on 2005 Feb 24, 2005 Jul 18 and 2005 Aug 14.  Observations
were also made on 2005 Jul 15, but the daytime phase stability was too
poor to permit reliable calibration of the data.  In all cases, 7 of
the 8 SMA antennas were available.  The observations on 2005 Feb 24
were made in good nighttime winter conditions, with optical depth
towards zenith at
225\,GHz $\tau_{225}\sim$0.04 and 10\% humidity.  The summer
observations suffered somewhat from afternoon atmospheric turbulence,
and were made with $\tau_{225}\sim$0.05 and 40\% humidity on 2005 Jul
18, and $\tau_{225}\sim$0.12 and 20\% humidity on 2005 Aug 14.

The SIS receivers were tuned to a center frequency of 345.796~GHz in
the upper sideband for 2005 Feb 24, and 230.538~GHz in the upper
sideband for 2005 Jul 18 and 2005 Aug 14.  For the initial observations
on 2005 Feb 24, one IF was configured with a higher spectral resolution
to search for CO(3-2) at the systemic velocity of M81, but none was
detected.  This is consistent with the absence of CO(1-0) emission at
the position of the core reported by \cite{Sakamotoetal2001}.  For all
observations, the full 4 GHz (2 GHz in each sideband separated by 10
GHz) were averaged to construct one continuum channel centered on
340.67 GHz and 225.42 GHz, respectively.

The SMA data were calibrated using the MIR software package developed
at Caltech and modified for the SMA.  Gain calibration was performed
using the nearby quasar 0958+655.  Absolute flux calibration was
performed using Callisto, and at least 2 of the following quasars--
0721+713, 0841+708, 0927+390, 1153+495-- were observed hourly to
ensure that any detected changes in the flux of M81* were real, and
not an artifact of the calibration.  This is particularly important
for the summer observations, which were made under poorer daytime
conditions.  The data were imaged using {\tt difmap} to confirm that
M81* is unresolved at the 1$\farcs$5-3$\farcs$0 resolution of the SMA,
following which the fluxes of both M81* and the nearby quasars were
determined by fitting a point source model to the data in the {\it
${u,v}$} plane.  The flux densities obtained are accurate to within
20\%, based on the derived values for the quasars.

As shown in Figs.~\ref{smafeb}--\ref{smaaug}, M81* exhibited
  little to no variation on short timescales during the observations,
  although there may have been a brief dip in flux in the February
  data shortly preceding the apparent corresponding dip in the PdBI
  measurements shown in Fig.~\ref{febpdbsma}.  This dip should be
  viewed with caution, given that the level of variation is not
  greater than the random variations seen in the calibrators, however
  the shape and timing are suggestive.  More significant variations in
  flux were seen between the epochs, as shown in Fig.~\ref{1mmpdbsma}.

The average flux on 2005 Feb 24 was 378.7$\pm$70.0 mJy at 340 GHz. For
  2005 Jul 18 and 2005 Aug 14, we obtained 182.8$\pm$36.0 mJy and
  91.5$\pm$15.3 mJy at 225 GHz, but it should be noted that the August
  data were obtained in substantially worse weather, with higher
  atmospheric opacity in addition to the normal summer daytime
  turbulence, and so the phase transfer from 0958+655, although only 4
  degrees away, may not have been entirely successful, resulting in a
  lower flux density within a point source model.

\subsection{IR: Lick Observatory} \label{subsec:lick}

We observed M81$^*$ with the infrared camera for adaptive optics at
Lick (IRCAL) behind the laser-guide star adaptive optics (LGSAO)
system on the Shane 3 m telescope on the nights of 2005 February 24
and 25 (see Tab.~\ref{febtimes} and Figs.~\ref{campoview} and
\ref{febview}). In the near-infrared, M81$^*$ is an unresolved point
source on top of a bright, extended background from the stars and gas
in the nuclear regions.  The high galactic background made use of the
laser guide star necessary; attempts to use the bright nucleus as a
natural guide star for AO correction failed.

We cycled through observations in the $J$, $H$, and $K_{\rm s}$ bands,
with more time spent on the $K_{\rm s}$ band where the correction is
best. For each cycle a 5-point dither pattern was repeated twice.
Each frame in the dither pattern was 50--90 s long in the $J$ band, 60
s long in the $H$ band, and 60 s long in the $K_s$ band.  We created
flats by taking the median of a series of images of the telescope dome
with the lights turned on, and divided each frame by the flat.  We
then subtracted the sky background from each frame using a nearby,
relatively blank field.  The sky background was small compared to
the bright extended emission from the galaxy.

Co-added images were created by shifting and adding the images from
each 5-point observing pattern. The shifts between the dithered images
were empirically determined between frames using the centroid of a
2-dimensional Gaussian fitted to the emission from M81$^*$ before
summing the frames.  Frames with an average $\sigma>11.2$ pixels
(0\farcs85) were excluded; this cutoff value was determined
empirically by plotting the peak pixel value versus the average
$\sigma$ for each frame.  (This average $\sigma$ value is from a
single Gaussian fit to the extended galaxy plus M81$^{*}$.) Co-added
images were then divided by an exposure map with the total exposure
time per pixel to create fluxed images in units of counts~sec$^{-1}$.

Each night we obtained two images of a nearby star (GSC 04383--00224)
in order to test the stability of the point-spread function (PSF), and
these images revealed the PSF to be variable. For instance, in the
$K_s$ band, the best PSF image was nearly diffraction-limited with a
half-width half-maximum (HWHM) of 0\farcs17, but the poorest on the
same night had a HWHM of 0\farcs25. No other point source was detected
in the 20\arcsec\ field of view of M81$^*$ that would allow us to
track the shape of the PSF concurrent with each exposure, and we found
that without empirical knowledge of the PSF we were unable to
construct a self-consistent model of the light from the galaxy and the
active nucleus. This prevented us from obtaining accurate point-source
photometry of the nucleus, and thus constraining any variability.
Therefore, we computed an upper limit to the $K_s$ band nuclear flux
averaged over the course of the observations.

\begin{figure*}
\epsscale{1}
\plotone{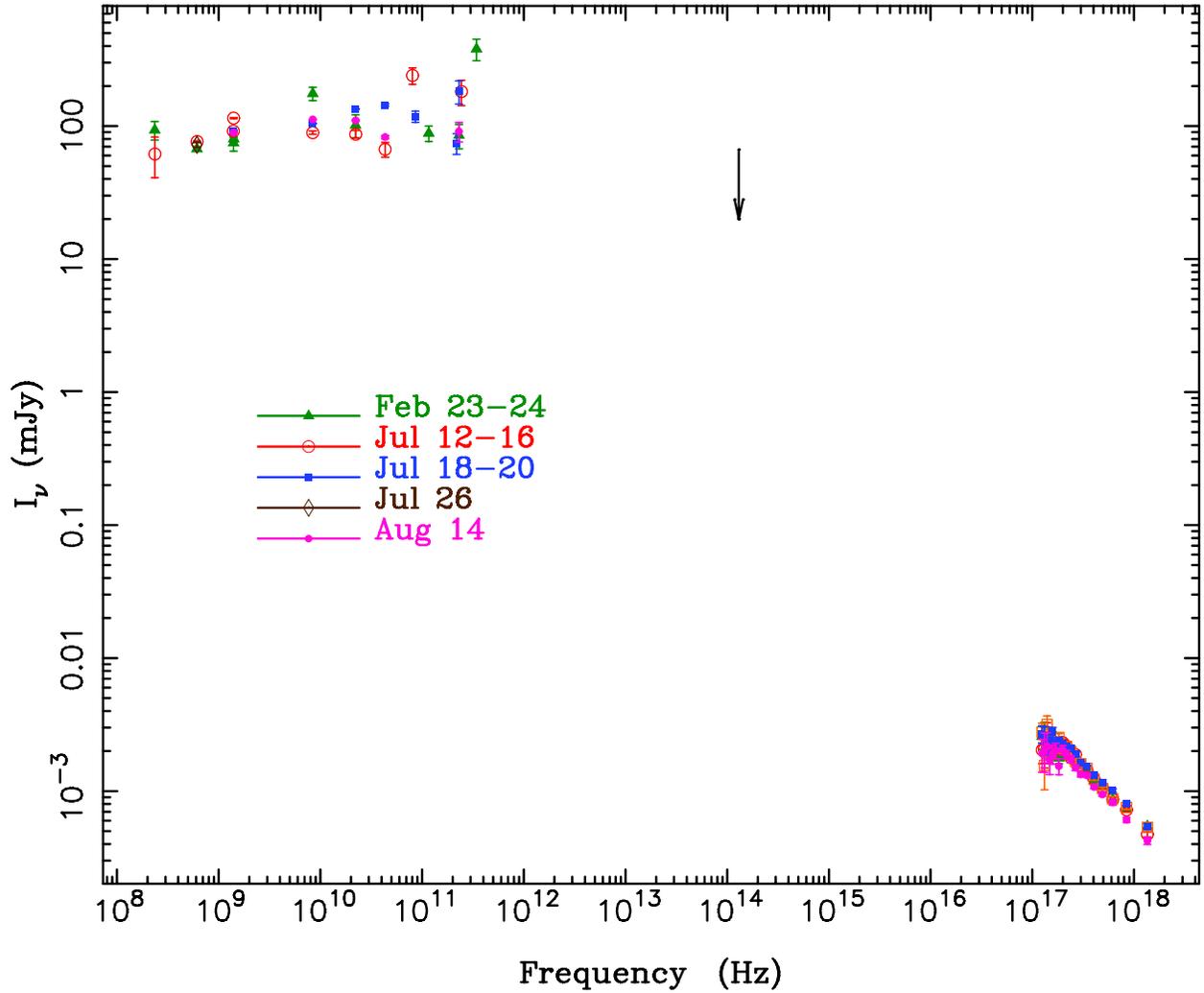}
\caption{The combined spectral energy distribution from all
  observations in the 2005 campaign.  (Dates are labeled in different
  colors for the electronic version.)  The base of the arrow
  represents the upper limit established by Lick, while all other
  points represent actual detections with estimated errors.  X-ray
  data are for the {\em MEG} $\pm 1^{\rm st}$ order spectra
  (combined). \label{alldata_nouls}}
\end{figure*}

To determine the upper limit, we measured the radial profile of the
PSF calibration star, GSC 04383--00224, and used its 2MASS $K_{\rm s}$
magnitude to determine the photometric zeropoint for converting from
counts to magnitudes using a large aperture (30 pixels = 2\farcs28).
At radii larger than 30 pixels, the background-subtracted, enclosed
counts change by $<2\%$ while the noise increases.  Given that M81$^*$
lies on top of strong extended emission from the galaxy, we preferred
to use a smaller aperture of 10 pixels (0\farcs76) with the aperture
correction determined from the calibration star; a 10-pixel radius
circle encloses 79\% of the counts within a 30-pixel aperture.  The
local background was estimated from the median in an annulus with
inner radius of 30 pixels and outer radius of 40 pixels (3\farcs04) and
subtracted; this likely underestimates the true local background as the
extended galaxy emission is also peaked at the position of M81$^{*}$.
The calibration star photometry, measured at three different airmasses
(1.20, 1.32, and 1.36) was also used to calculate the local $K_{\rm
s}$-band atmospheric extinction correction. Following this procedure,
we found an upper limit value of 66.7 mJy in $K_{\rm s}$.  This limit
is displayed as the base of the arrow in Figure ~\ref{alldata_nouls}.

\subsection{X-rays: \textit{Chandra}} \label{subsec:chandra}

M81* was observed by the \emph{Chandra} X-ray observatory, with the
High-Energy Transmission Grating Spectrometer \citep[{\em
HETGS};][]{Canizaresetal2005} in place, on five separate occasions
(see Tables~1--5).  The {\em HETGS} consists of two sets of
transmission gratings, the {\em High Energy Gratings} ({\em HEG})
covering the 0.8--10 keV bandpass with a spectral resolution of
$\Delta \lambda = 0.012$\AA\ FWHM, and the {\em Medium Energy
Gratings} ({\em MEG}) covering the 0.4--8.0 keV bandpass with a
spectral resolution of $\Delta \lambda = 0.023$\AA\ FWHM.  The angular
resolution of \emph{Chandra}, even with the insertion of the gratings,
isolates X-ray emission from the central $< 1\arcsec$ around M81*.  We
do not utilize information from the $0^{\rm th}$ order (undispersed)
spectrum, however, as the central image of the nucleus suffers from
photon pileup \citep{Youngetal2007}.

The \emph{Chandra} data were filtered for times of high background and
spectra were extracted using the standard CIAO
tools\footnote{\url{http://cxc.harvard.edu/ciao/}}, using v3.3 of the
software and CALDB 3.2.2. Analyses were performed using ISIS version
1.4.7 \citep{Houck2002}.  Our data preparation was identical
to that described in \citet{Youngetal2007}.  Specifically, for all
X-ray analyses we separately combined the $\pm1^{\rm st}$ {\em HEG}
spectra and the $\pm 1^{\rm st}$ order {\em MEG} spectra, and we
utilized background files from narrow regions on either side of the
respective gratings arms.  As discussed in \citet{Youngetal2007}, we
are confident that $\aproxgt 90\%$ of the dispersed X-ray emission
originates in an unresolved source in the nucleus of M81.

\section{Observational Results}\label{sec:newobs}

\subsection{The Broadband Spectrum of M81*}\label{subsec:broad}

Figure~\ref{alldata_nouls} shows the broadband spectrum comprised of
all of our observations from the 2005 campaign.  It is immediately
clear that while there was some variability, particularly in the radio
to mm, we did not see large variations in the X-ray or overall basic
shape of the SED.  The total average 7 GHz radio luminosity is $L_{\rm
  R}=8.2\times10^{36}$ erg s$^{-1}$ and the total average 2--10 keV
X-ray luminosity is $L_{\rm X}=1.52\times10^{40}$ erg s$^{-1}$.  The
radio exhibited $\approx20$\% variation about this average while the
X-rays exhibited $\approx14$\% variation.  Therefore over the course
of this half-year campaign, M81* appeared to be more stable in both
wavebands than has been reported in the past.  Interestingly, however,
the average $L_{\rm R}$ seen in our campaign is $\approx25$\% lower
than the average deduced from all prior campaigns, while our average
$L_{\rm X}$ is about a factor of 5 less (see \citealt{Markoff2005}).
Either we have caught M81* in a rather low, stable state, or previous
observations (with notably larger fields of view) have included a large
flux contribution from the surrounding diffuse medium.

\begin{figure*}
\epsscale{1} \plotone{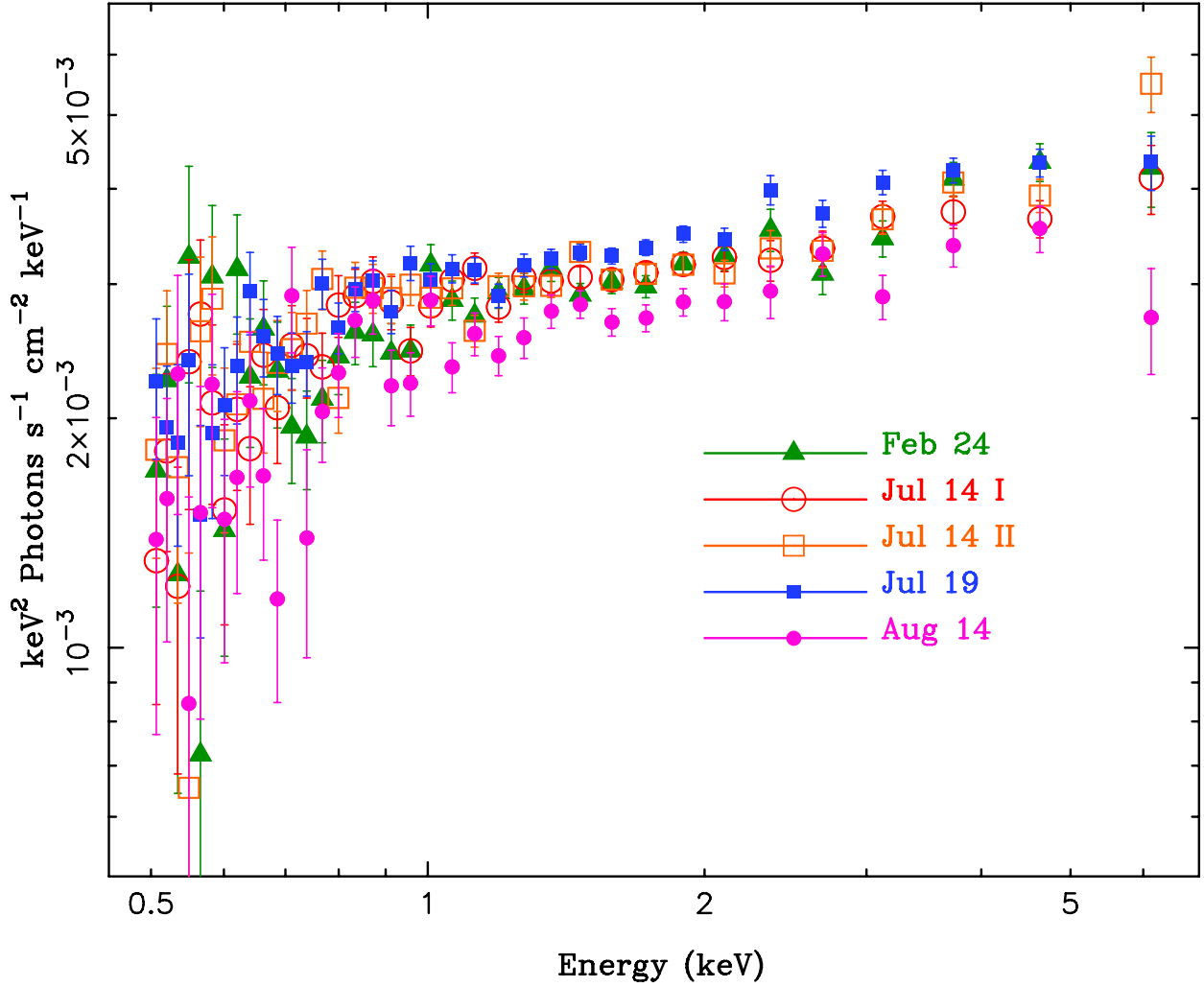}
\caption{Flux density ($\nu L_\nu$) of the X-ray band for all
  observations during the campaign (in color for the electronic
  version).  The slope for all observations is well-fit by a power-law
  of photon index $\Gamma \approx 1.75$--1.85.  The largest change in
  continuum was between July 19 and August 14, with a drop of 
  $\approx 20$\% within a month.
  \label{nulnuxrays}}
\end{figure*}

Fig.~\ref{nulnuxrays} shows the flux density for all X-ray
observations, where we see that the largest change was a drop between
July and August.  The 2005 July 19 observations with the {\em HEG}
showed a 0.8-7\,keV flux of $1.31\pm0.02\times10^{-11}~{\rm
erg~cm^{-2}~s^{-1}}$--- the highest value during our campaign--- while
the 2005 August 14 observations with the {\em HEG} dropped down to a
value of $1.07\pm0.03\times10^{-11}~{\rm erg~cm^{-2}~s^{-1}}$.  (See
\citealt{Youngetal2007} for further discussions of the X-ray
variability.)

\begin{figure*}
\epsscale{1}
\plotone{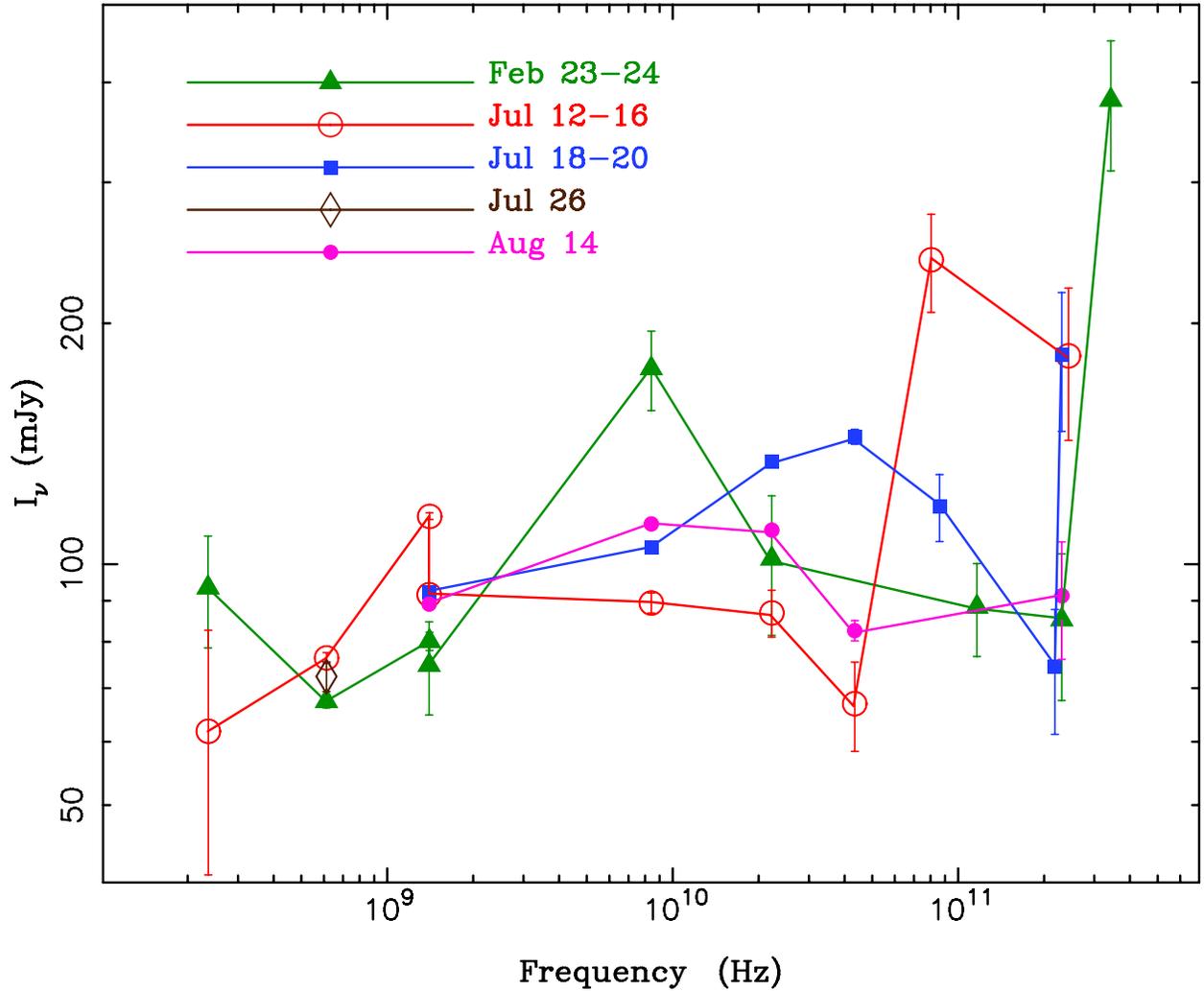}
\caption{Spectral energy distribution of the centimeter through submm
  observations from the campaign (in color in the electronic version).
  Note the significant variability, especially at the higher
  frequencies.  We detect significant intraday variability during both
  2005 February and July, as well as apparent waves moving lower in
  frequency and amplitude consistent with adiabatic expansion of blobs
  in a jet. \label{radio}}
\end{figure*}

\begin{figure*}
\includegraphics[width=\textwidth]{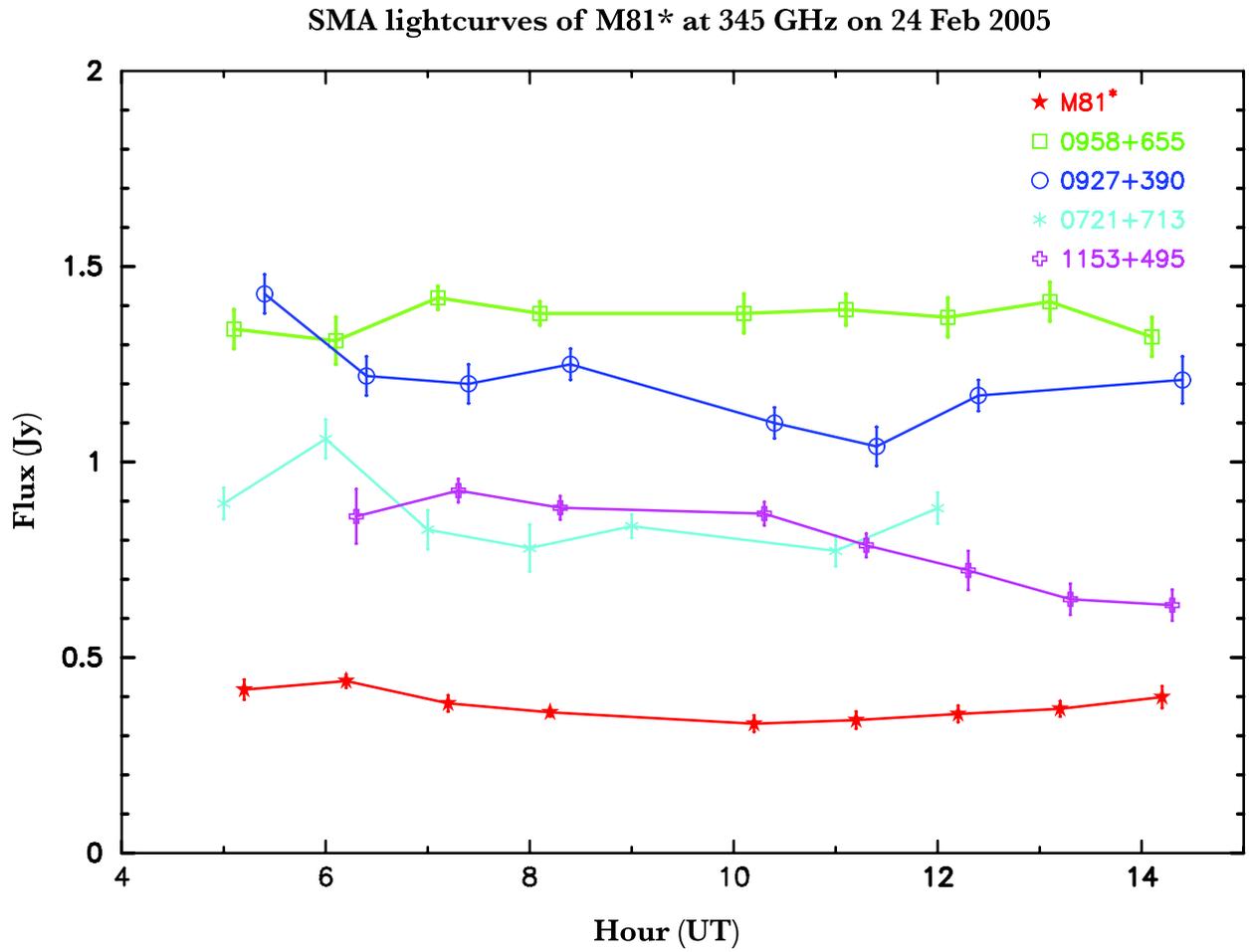}
\caption{Lightcurve of 24 Feb SMA observations of M81* and
  calibrators.   \label{smafeb}}
\end{figure*}

\begin{figure*}
\includegraphics[width=\textwidth]{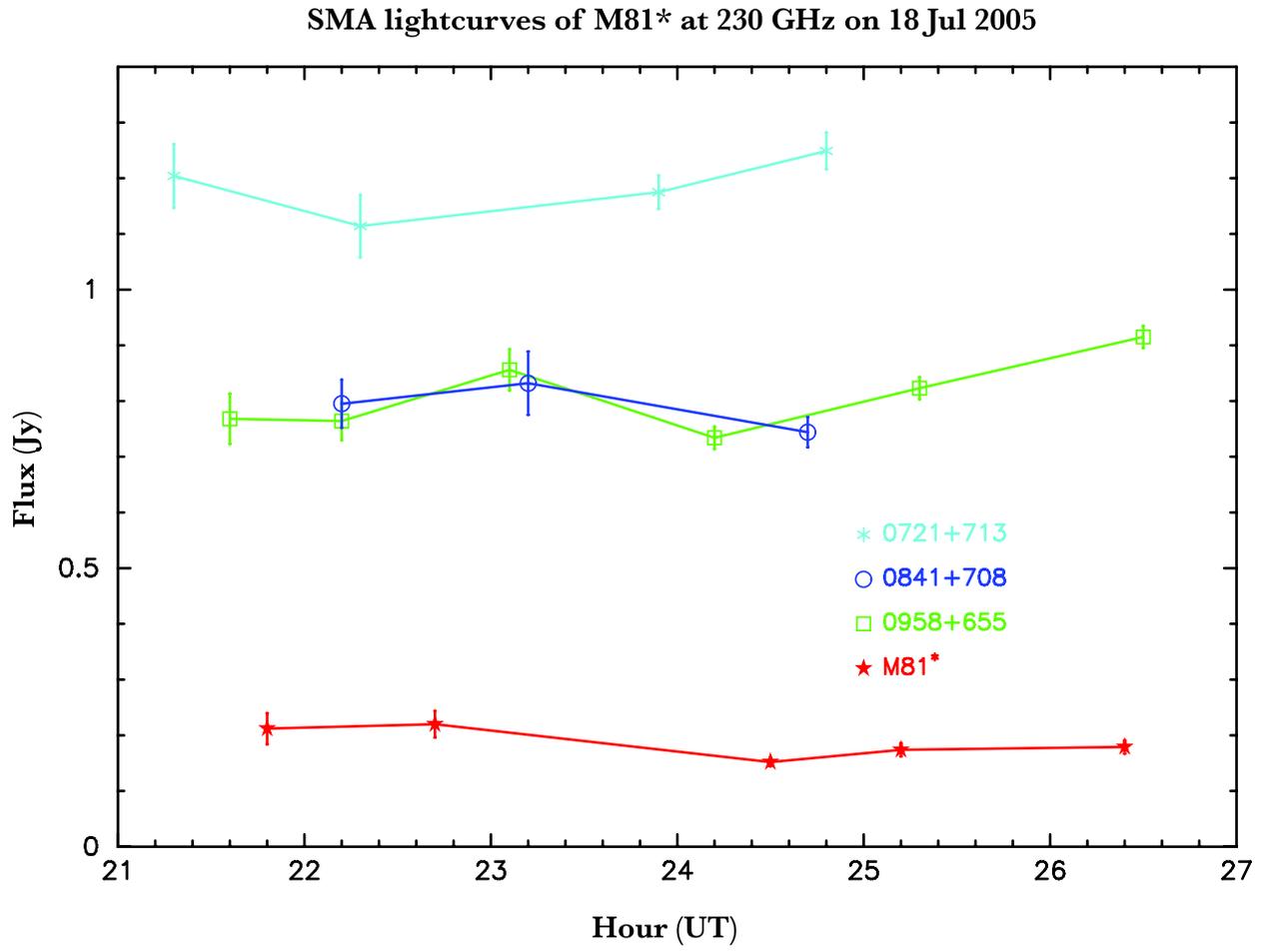}
\caption{Lightcurve of 18 July SMA observations of M81* and
  calibrators.   \label{smajul}}
\end{figure*}

\begin{figure*}
\includegraphics[width=\textwidth]{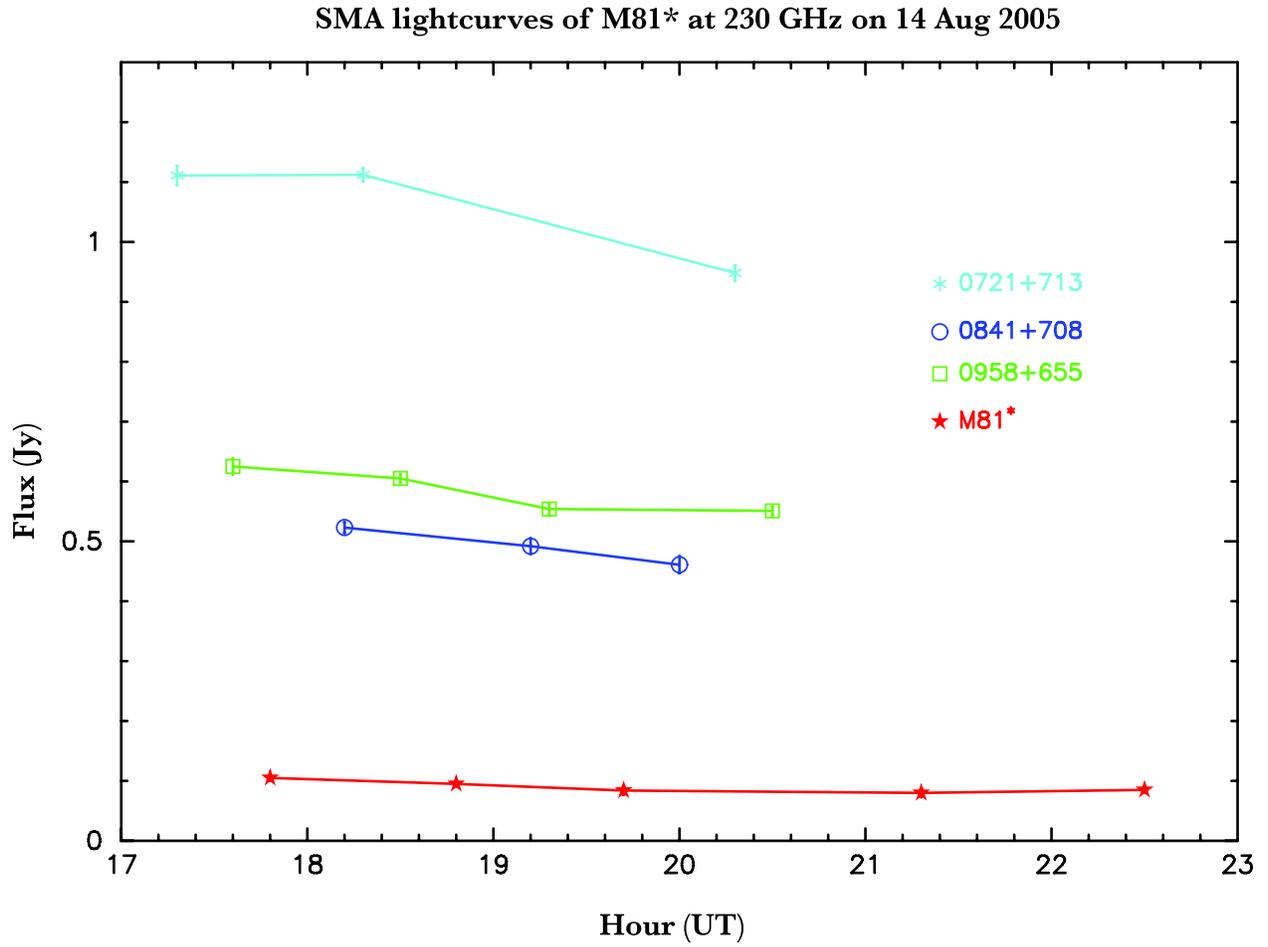}
\caption{Lightcurve of 14 Aug SMA observations of M81* and
  calibrators.   \label{smaaug}}
\end{figure*}

\begin{figure*}
\includegraphics[width=\textwidth]{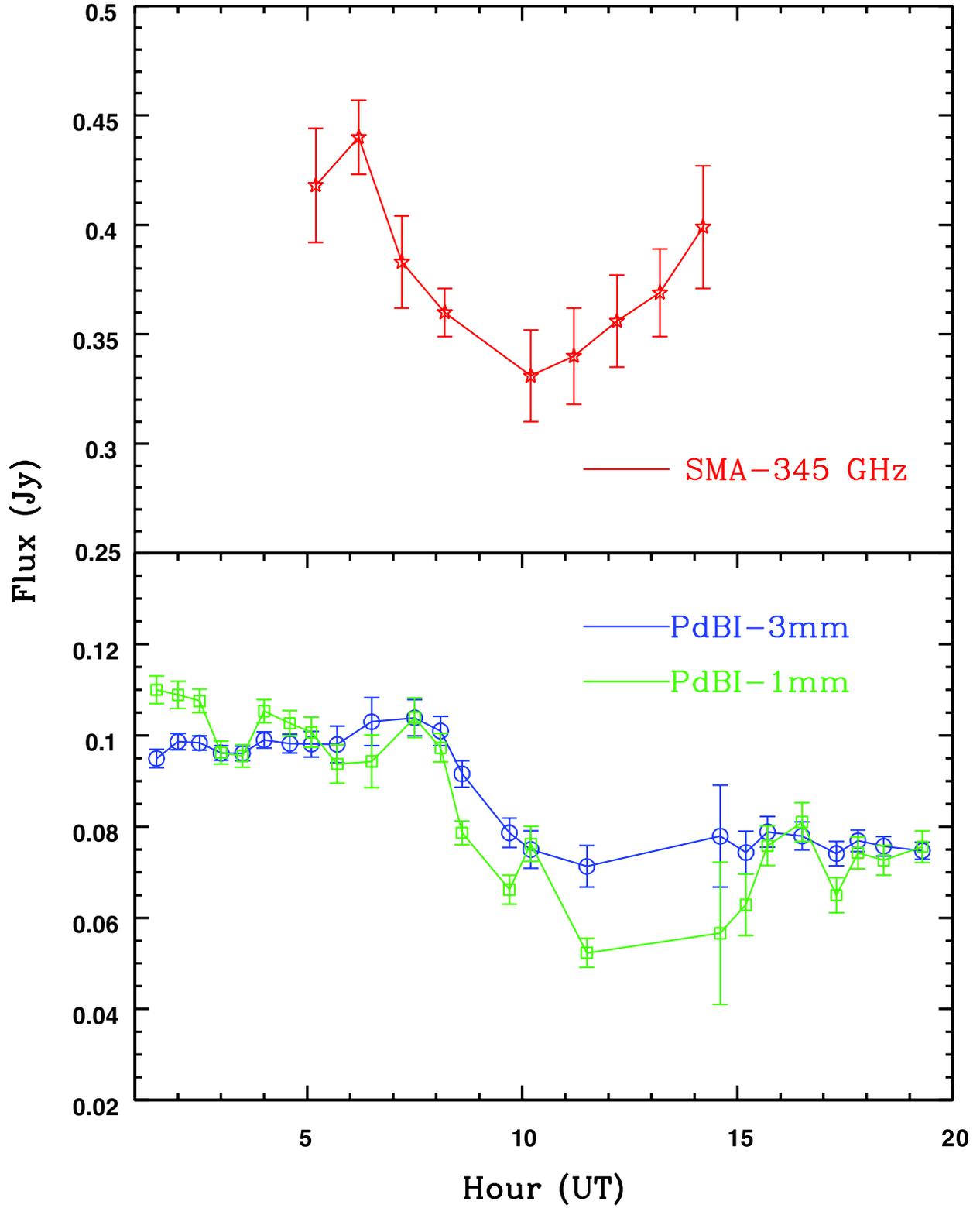}
\caption{Comparison of PdBI and SMA lightcurves from 24 Feb.   
\label{febpdbsma}}
\end{figure*}

The radio through submm bands revealed significantly more variability
in comparison to the X-ray band.  Fig.~\ref{radio} shows the good
data for all observations.  The most reliable detection of significant
intraday variability occurred at 3mm and 1mm in the PdBI observations
on 24 February.  A flux decrease of $\sim30$\% with a significance of
over $5\sigma$ was observed in both bands from 08h to 12h UT (see
Fig.~\ref{febpdbsma} and \citealt{Schoedeletal2007}).  Such a time
scale would suggest that the size of the emitting region is less than
20\,$r_{\rm g}$, if beaming were not involved.  The expected beaming
from our spectral fits discussed below is mild for the weak jet in
this LLAGN; therefore, the mm variability still implies a relatively
compact source.

The SMA observations were successful in the 345 GHz band only on the
first observing run in February.  The resulting spectrum shows a
suggestive upturn towards the submm (Fig.~\ref{radio}, \ref{febfit}).
This steepening is similar to the spectral component seen in Sgr~A*
that is referred to as the ``submm bump''.  In Sgr~A*, this bump
rapidly declines with decreasing wavelength towards the infrared (IR),
and furthermore it varies simultaneously with the X-rays
\citep{Eckartetal2004,Eckartetal2006a}. This simultaneous variation
suggests that for Sgr~A* the IR and X-ray emission both originate in
regions close to the SMBH.  In contrast, the variability detected in
M81* is both less pronounced and not as clearly correlated between the
submm and X-ray.

That being said, however, the low-frequency data are suggestive of
waves of variability, with decreasing amplitudes, that appear to be
moving from shorter to longer wavelengths over the half year of
monitoring.  Specifically, note that the peak at just under 10 GHz in
2005 February is gone by 2005 August, and does not appear to be
associated with any lower frequency features. (The ``peak'' in the
2005 July 12-16 at just above 1 GHz is a mismatch between the GMRT and
the VLA, which may be due to real intraday variability as there was
$\sim 29$ hours between the two observations).  One question that
arises is whether or not the doubling of flux density at 43\,GHz that
occurs between 2005 July 13 and 19 is associated with the 2005 July 14
peak at 80.5\,GHz moving to lower frequency over the ensuing several
days.  Likewise, is the 43\,GHz peaked bump on 19 July then seen moved
to lower frequency (10--20\,GHz) and amplitude in the 2005 August 14
data?  Such a ``wave of variability'' is consistent with expectations
of adiabatic expansion.  Similarly, the February observation with
contiguous PdBI and SMA observations (Fig.~\ref{febpdbsma}) shows a
clear dependence of variability amplitude on frequency, which is
another expectation of adiabatic expansion.

\subsection{Comparison to Sgr~A*}

Our new simultaneous data reinforce the similarities previously
reported between M81* and Sgr A*.  In Fig.~\ref{m81sgra} we show the
total M81* campaign spectrum, including the non-simultaneous IR/O/UV
data upper limits discussed in \S\ref{subsec:uv}, and overplot the
simultaneous Sgr~A* spectrum from \cite{Anetal2005} along with the
various {\em Chandra} X-ray spectra
\citep{Baganoffetal2001,Baganoff2003,Baganoffetal2003}.  The Sgr~A*
data have been scaled downward by a factor of $\approx10$ in order to
ease visual comparison.

M81* and Sgr A* show remarkable similarities in the radio frequencies.
Even though for M81*, within the GMRT error bars, we see no direct
evidence for the free-free absorption turnover observed in Sgr A*, the
slightly inverted spectra in both sources are still classic indicators
of synchrotron self-absorption effects in the jet core emission.  The
M81* radio spectrum lies below that of Sgr~A* at higher frequencies,
and thus is less inverted than that of Sgr A*. This is expected for
M81* in the context of self-absorbed, accelerating jet models as its
lower inclination angle compared to Sgr~A* would result in a less
inverted spectrum.  The jets in Sgr A* are presumed to be at a high
inclination angle with respect to our line of sight (see the
supporting evidence from the models presented by
\citealt{MarkoffBowerFalcke2007,Meyeretal2007}).  The spectra for M81*
and Sgr~A* both seem to peak near the submm range and then
subsequently drop off towards the IR. Both sources also are consistent
with sharing the same radio-IR power-law slope, if the M81* IR
upper-limits are indicative of the underlying intrinsic spectrum.
M81* and Sgr~A*, however, clearly diverge from one another in the
X-ray regime.  On a relative scale, the M81* X-ray spectrum always lies
above that for Sgr A*, with the latter source in its rare, bright
flare states still falling short of the X-ray/radio flux ratio of
M81*.

If M81* indeed has a ``submm bump'' as in Sgr A*, this would be only
the second source where such a feature is observed.  In Sgr A* this
component has been associated alternatively with the base of a compact
jet \citep{FalckeMarkoff2000,Markoffetal2001}, the regions of the
accretion flow closest to the black hole
\citep{Narayanetal1998,YuanQuataertNarayan2003}, or with an inner
Keplerian disk \citep{LiuMelia2001}.  Regardless of the exact
geometry, in Sgr A* this component now has been definitively
associated with X-ray flares \citep{Eckartetal2004,Eckartetal2006b},
and thus is of significant interest for understanding high-energy
processes within tens of $r_g$ of the Sgr A* black hole.  For the
LLAGN class as a whole, it is important to understand if the flaring
and coupling of the submm bump/X-ray emission in Sgr A* is typical.
The X-rays in M81* have yet to show any flares of significant
amplitude, while we have some evidence of variability in the submm.
Thus in contrast to Sgr~A*, the X-ray emission in M81* may not be due
to the same physical component that yields the variable submm.  As
further described below, this has implications for the theoretical
modeling of M81*.

These results for the submm band in particular are still tentative;
more monitoring of this band as well as the millimeter range will
determine whether an intrinsic submm bump is indicated.  Massive
amounts of dust are present in the region observed in these bands and
could be contributing some level of flux.  On the other hand,
subsequent brief observations at 345 GHz using the SMA in 2005
(A. Peck, priv. comm.) show significant variability, yielding a flux
density ranging from 300 mJy to 900 mJy over a period of 3 months
from April to June 2006, consistent with the measurements
made at the SMA in Feb 2005.  These values are accurate to
within about 20\%, and thus the variability is clearly significant.  
These further data also support a rise in flux above the millimeter
band.  It is possible that the submm bump is a completely transient
feature associated with flaring/ejecta, as has been suggested for Sgr
A*.  Confirming this strong variability as well as a detection of
linear polarization would place stringent limits on any dust
contribution, as well as the spectrum.  With the advent of
polarimetry at 345 GHz this year on the SMA, we have (with D. Marrone,
PI) successfully proposed to search for linear polarization in M81*, which
will hopefully resolve this issue in the near future.

\section{Jet-dominated spectral models}\label{sec:model}

\subsection{Model description}\label{subsec:jet_model}

One of the basic tenets of General Relativity is that black holes are
essentially self-similar with regards to mass.  An obvious possible
consequence of this is a predictable scaling with mass of the
accretion physics around black holes.  If such a scaling exists, it
would imply that the same underlying physical model could explain the
continuum emission for stellar accreting black holes in XRBs as well
as SMBHs.  Likely complicating this simple picture are differences
that could be introduced by accretion off one star compared to
accretion off the winds of entire clusters of stars.  Furthermore XRBs
undergo state changes, which have yet to be clearly associated with
the various AGN classes as would be naively expected if some of the
AGN classes correspond to much longer lived state transitions.

So far the best case for such a mapping, referred to as the
``fundamental plane of black hole accretion'', is the correspondence
in characteristics between the sub-Eddington low/hard state of XRBs
and LLAGN
\citep{MerloniHeinzDiMatteo2003,FalckeKoerdingMarkoff2004,KoerdingFalckeCorbel2006}.
The hard state of XRBs is characterized by weak accretion disk
emission, and the presence of steady, compact jets
\citep[e.g.][]{Fender2006}.  These jets seem to increasingly dominate
the power output of the system even as the total accretion rate
decreases \citep{FenderGalloJonker2003}.  Lending theoretical support
to the idea of a fundamental plane, models of LLAGN as being dominated
by outflows
\citep[e.g.][]{FalckeMarkoff2000,Markoffetal2001,Yuanetal2002} also
have been quite successful at explaining the broad continuum
properties of hard state XRBs
\citep[e.g][]{MarkoffFalckeFender2001,Markoffetal2003}.  More
recently, these outflow-dominated models have been refined further to
account for strong geometrical constraints, for example, such as
signatures of reflection off of cool material.  The outflow models
were developed to take advantage of data from recent simultaneous,
multi-wavelength monitoring campaigns.  Statistical fitting of the
simultaneous broadband continua of several hard state XRBs, including
finer features such as fluorescent Fe lines in their X-ray bands, have
produced consistent trends among the physical parameters determined
for this class of sources \citep[][; Maitra et al., in
prep.]{MarkoffNowakWilms2005,Migliarietal2007,Galloetal2007}.

In order to further test the principle of black hole accretion scaling
with mass, and to enable a stronger physical comparison among M81*,
Sgr A*, and their possible stellar-mass equivalents, we apply these
same hard state XRB outflow-dominated models to the M81* spectra from
our campaign.  The only significant difference from the model
application to XRBs is that the input mass for M81* is $\sim10^6$
times larger.  A more detailed description of this model can be found
in the appendix of \cite{MarkoffNowakWilms2005}, and
references therein; here we provide a brief summary.

The model was designed to test the premise that the magnetized,
outflowing compact accretion disk coronae such as described in
\cite[e.g.][]{Beloborodov1999,MalzacBeloborodovPoutanen2001,MerloniFabian2002}
can comprise the footpoints of collimated jets.  Although
  magnetic fields are assumed to play a global role, the dynamics of
  the model do not include magnetic acceleration.  There are two
  primary reasons for this choice.  First, because the exact role of
  the fields in the dynamics is still under debate, including a
  magnetic pressure term explicitly would add more assumptions (and
  thus free parameters) to the model.  Secondly, the observations
  suggest that the steady jets in the weakly accreting black hole
  state are less accelerated than the transient jets occurring near
  the Eddington limit \citep[e.g.][]{Fender2006}.  The observations so
  far can thus be well-explained by a gas pressure dominated model.
  In addition, for M81* and Sgr A* in particular the inverted radio
  spectrum is also suggestive of adiabatic cooling in a jet with a
  broader opening angle ($\ge 20^\circ$), rather than a narrow jet as
  would be expected for magnetic collimation.

Aside from these points, there are four basic assumptions in
the model: 1) the total power in the jets scales with the total
accretion power at the innermost part of the accretion disk,
$\dot{M}c^2$, 2) the jets are freely expanding and only weakly
accelerated via their own internal pressure gradients, 3) the jets
contain cold protons which carry most of the kinetic energy, while
leptons dominate the radiation, and 4) some or all of the originally
thermally distributed particles are accelerated into a power-law which
is maintained along the rest of the jet via distributed acceleration.

The base of the jet consists of a small nozzle of constant radius
where no bulk acceleration occurs.  The nozzle absorbs our
uncertainties about the exact nature of the relationship between the
accretion flow and the jet, and fixes the initial value of most
parameters.  Beyond the nozzle, the jet expands laterally with its
initial proper sound speed for a relativistic electron/proton plasma,
$\gamma_{\rm s}\beta_{\rm s}c\sim0.4c$.  The plasma is weakly
accelerated by the resulting longitudinal pressure gradient force,
allowing an exact solution for the velocity profile via the Euler
equation \citep[see][]{Falcke1996}.  This results in a roughly
logarithmic dependence of velocity upon distance from the nozzle, $z$.
The velocity eventually saturates at large distances at Lorentz
factors of $\Gamma_{\rm j}\ga$2-3.  The size of the base of the jet,
$r_0$, is a free parameter and once fixed determines the radius as a
function of distance along the jet, $r(z)$.  There is no radial
dependence in this model, and the opening angle is thus fixed
  by the velocity profile as a function of distance along the jets.

The model is most sensitive to the fitted parameter $N_{\rm j}$, which
acts as a normalization.  It dictates the power initially divided
between the particles and magnetic field at the base of the jet, and
is expressed in terms of a fraction of $L_{\rm Edd}$.  Once $N_{\rm
j}$ and $r_0$ are specified and conservation is assumed, the
macroscopic physical parameters along the jet are determined.  We
assume that the jet power is roughly shared between the internal and
external pressures.  The radiating particles enter the base of the jet
where the bulk velocities are lowest, with a quasi-thermal
distribution of temperature $T_e$ (a fitted parameter).  A significant
fraction (here fixed at 75\% based on results from the XRB fits
mentioned above) of the particles are accelerated into a power-law
tail at a location $z_{acc}$ (also a fitted parameter).  The maximum
energy the electrons can achieve is calculated explicitly by setting
the local cooling and escape rates to the acceleration rate.  Here we
assume that the particles are accelerated via the most conservative
case of diffusive shock acceleration with the magnetic field parallel
to the shock normal \citep[see][]{Jokipii1987}.  The acceleration rate
depends on the plasma parameters $u_{\rm sh}$ and $\xi$, the relative
velocity of the shock to the plasma in the shock frame, and the ratio
between the mean free path for scattering and the gyroradius,
respectively.  These terms enter into the acceleration rate as
$\epsilon_{\rm sc}\equiv\left(u_{\rm sh}/c\right)^2/\xi$.  This is a
free parameter in our fits, and since $\xi$ is thought to lie in the
range $\sim 10$--$1000$, it provides a consistency check on the plasma
velocities.

The particles in the jet radiatively cool via adiabatic expansion,
synchrotron processes, and inverse Compton upscattering; however,
adiabatic expansion is assumed to dominate the observed effects of
cooling.  While thermal photons from the accretion disk (via a
multicolor blackbody model) are included as seed photons in the
inverse Compton calculations, beaming reduces their energy density
compared to the rest frame synchrotron photons (synchrotron
self-Compton; SSC), except at the very base of the jet where they can
be of the same order.  The temperature $T_{\rm d}$ and total
luminosity $L_{\rm d}$ of a thin accretion disk are also included as
fitted parameters, but as this component is not an integral part of
the outflow model, and not well constrained by this data set, we
include these parameters mainly for a consistency check.

Aside from those mentioned above, the other main fitted parameters are
the ratio of nozzle length to its radius, $h_0$, and the equipartition
parameter between the magnetic field and the radiating (lepton)
particle energy densities, $k$.  Physical parameters such as the mass,
inclination angle and distance are fixed at the observed values.
Table~\ref{tab:params} summarizes all the fitted jet parameters.

This model has provided a very good statistical description of the
broadband radio through X-ray data from several XRBs.  The data from
this campaign observing M81* allows us to test the premise that
sub-Eddington accretion in LLAGN can be described by the same physics
as sub-Eddington accretion in low/hard state XRBs, even though the
latter sources are over six orders of magnitude less massive than the
former.  This is the first time that our jet model has been applied to
a ``canonical'' LLAGN, which thus allows for an interesting comparison
to Sgr A*.

\subsection{Fitting methodology}\label{sec:fitting}

We explore two possible jet-model scenarios with our M81* data.  The
first scenario is an exact analog to the XRB fits, where the particles
are presumed to enter the nozzle in a quasi-thermal distribution that
is later accelerated in the jet.  This scenario was also explored for
Sgr A*, where we determined that the acceleration must be either
extremely weak or lacking, or that it must occur extremely far out in
the jets, in order to prevent the presence of a predicted power-law
violating the observed IR flux values \citep{MarkoffBowerFalcke2007}.
The second scenario assumes that the particles enter the jets in a full
power-law distribution, as may occur if jet formation is associated
with a shock in the inner regions of the accretion flow
\citep{Koideetal2000}. This scenario was also explored for Sgr A*,
where in order to be consistent with the quiescent spectrum the acceleration
must be sufficiently weak to result in an electron particle distribution
power-law no harder than $p\sim3.8$ \citep{MarkoffBowerFalcke2007}.

In exploring these models, fits to the individual ObsIds for the {\em
Chandra} data were simultaneously carried out over the 0.5--7\,keV
band for the {\em MEG} data and the 0.7--8\,keV band for the {\em HEG}
data.  Each dataset was grouped to have a minimum signal-to-noise of 5
and a minimum of four wavelength channels per bin.  Because there are
hundreds of X-ray data points compared to the radio/submm data, we
initially set the radio/submm data errors to $<1\%$ to ``weight'' the
significance of the sparser lower-frequency data in the fitting
procedure.  But because the jet model produces a predominantly smooth
radio spectrum (i.e., it does not account for the waves of variability
likely moving through the radio spectrum), and because the radio error
bars are for the most part small compared to the
observation-to-observation variability, in our final joint radio/X-ray
fits we add 20\% error bars in quadrature to the statistical error bars
for the radio data.  In this manner we decrease the likelihood that
the jet models fall into local minima, and ensure that they instead
more fairly represent the average radio properties. 

We also present a series of fits wherein we simultaneously consider
the radio and X-ray data from the entire campaign.  We again add 20\%
error bars in quadrature to the statistical error bars of the radio
data.  For the {\em Chandra} data we use the {\tt ISIS} {\tt
combine\_datasets} function\footnote{For our purposes here, this is
equivalent to adding the data via {\tt ftools} functions outside of
the fitting program; however, utilizing this function within {\tt
ISIS} allows more flexibility in modeling and plotting.}  to combine
the {\em HEG} data into a single spectrum and to combine the {\em MEG}
data into a single spectrum.  For each of these spectra we further
group the data such that there is a minimum of 100 counts and a
minimum of 32 energy channels per bin.

The addition of 20\% error bars to the radio and optical data
  acts to subsume both instrinsic variability, and allows for any
  systematic calibration differences among the different detectors.
  However, one might worry that the radio and optical data would then
  apply little statistical leverage on the fits.  In fact, owing to
  the 2.5 to 9 orders of magnitude energy differences between the
  X-ray and UV/radio data, their effect far outweighs the simple
  contribution calculated to the overall $\chi^2$.  For example, if
  one pivots a power-law with $\Delta \Gamma = 0.01$ at 1\,keV, the
  change in slope would yield well over a 20\% difference in the radio
  regime (i.e., greater than our added error bars) while giving only a
  2\% difference at 8\,keV (i.e., substantially less than the X-ray
  data bars).  The jet models are thus predominantly constrained by
  the radio data via this effect, rather than their contribution to
  the $\chi^2$ statistic.

In order to illustrate this explicitly, in
  Fig.~\ref{radioresids} we present the radio residuals for a fit
  performed solely in the X-ray regime, using jet model parameters
  somewhat different than the typical (based on fitting X-ray
  binaries) values discussed here. The model provides an extremely
  good fit to the X-ray data, with a change of $\Delta \chi^2 = 1.2$
  compared to our best fit from the broadband spectrum. This model,
  however, fails completely in the optical and radio regime, even with
  our expanded radio error bars.  Thus the simultaneous radio/submm
  data especially are crucial for excluding large regions of
  ``reasonable'' parameter space for the jet models.  The X-ray data
  alone simply cannot constrain jet physics.

The resulting parameter differences between our canonical fits
  and the two fits utilizing non-simultaneous upper limits in the
  infra-red and optical, further emphasizes that broadening the
  multi-wavelength coverage is clearly very desirable. The jet model
  fits do present a number of ``local minima'', which are also
  evidenced in the error bars for some parameter fits.  Occasionally
  we find large parameter error bars-- again an indication of inherent
  degeneracies in this complex theoretical model that we expect would
  be reduced with broader wavelength coverage.  We also occasionally
  find parameter error bars that are unusually small.  This is
  sometimes attributable to the fact that the radio data points do not
  represent a smoothed, averaged behavior, whereas the theoretical
  model does represent such an idealization to some extent.  If a fit
  fortuitously passes almost exactly through a few radio points, the
  fits occasionally become very localized to that fit local minima.
  Such a local minimum might be slightly removed from a separate local
  minimum associated with a different subset of the radio points.  The
  added 20\% radio error bars tends to reduce, but not entirely
  eliminate, this behavior of the fits.  Ultimately a time-dependent
  model is required to completely account for these effects.

For the X-ray band specifically, \citet{Youngetal2007} fit a series
of emission and absorption lines to the combined M81* {\em Chandra}
spectra.  To account for the presence of these features in the
spectra, for all of our fits we added the complete set of lines from
\citet{Youngetal2007}, but with there wavelengths frozen to those
found in that work, and their widths frozen to $10^{-4}$\,eV.  Only
the line amplitudes were allowed to vary in the fits.

\subsection{Results and interpretation}\label{sec:results}

Figs.~\ref{febfit} \& \ref{febfit_xray} present our best fits of each
type of model to the most comprehensive single observation in
February, while Figures~\ref{mongo} \& \ref{mongo_x} show the fits to
all campaign data combined.  Non-simultaneous data points from
\textsl{HST} \citep{Maozetal2005} have, however, been included
with error bars increased to 20\% in
order to guide the fits in the observational gap that otherwise
extends over six decades of photon frequency.  Figs.~\ref{mongo_ul} \&
\ref{mongo_ul_x} present additional fits which also include the upper
limits from the simultaneous Lick Observatory observations, as well as
several non-simultaneous detections taken from the literature (but
here used as upper limits since they likely include non-nuclear
emission) for ISO, MIRLIN, \textsl{HST} and \textsl{Spitzer}
\citep{Grossanetal2001,Gordonetal2004,Murphyetal2006}.  All of these
upper limit data were treated as measured points at their upper
values, with additional 20\% error bars applied.
Given the similarity in the shape of the submm through optical/UV data
from M81* compared to what is seen in Sgr A* and NGC~4258, it seemed
worthwhile investigating scenarios that included these limits as
estimates of the true, underlying spectra.

The values for the fitted parameters with 90\% confidence errors for
all fits, including the individual data set fits not shown here, are
presented in Tables ~\ref{tab:PL_FINAL_a}--~\ref{tab:MXSW_FINAL_b}.

It is clear that the model provides a very good description of the
individual observations, and to a lesser extent the combined data set.
The latter data set obviously includes variations, especially for the
non-X-ray data, for which the steady-state model cannot account.  What
is most striking about these fits is that the overall values for the
fitted parameters fall into remarkably similar ranges compared to the
parameter values for hard-state XRBs and Sgr A*.

We first consider the model with an initial power-law electron
distribution. The power-law electron distribution has previously been
applied to spectra from Sgr~A*, but not to any XRB spectra.  Compared
to the the range of Sgr~A* fits explored in
\cite{MarkoffBowerFalcke2007}, the fits to the M81* spectra a
similarly find a very compact jet base and an electron temperature of
$\approx 10^{11}$ K.  There are, however, significant differences
which are likely influenced by the lower X-ray/radio flux ratio in
Sgr~A*, as well as the fact that Sgr~A*, as a fraction of Eddington
luminosity, is over four orders of magnitude lower compared to M81*.
Whereas M81* shows indications of a weak accretion disk, no such
component has been observed in Sgr A*.  Whether assumed to occur near
the base, as for the power-law (PL) fits, or further out in the jets,
as in the Maxwellian (MXW) fits, the M81* spectra prefer solutions
with more efficient particle acceleration.  That is, the electron
power-law index $p=2.4$--2.8, whereas for Sgr~A* $p$ is always $>3$.
The M81* particle distribution also shows more cooling than in Sgr~A*,
as indicated by the $\sim 10$ times smaller $\gamma_{\rm max}$ in
M81*.  These differences go far to explain why the weak jets of M81*
have been easier to discern than those in Sgr A*.  Beyond the fact
that we are not viewing M81* through the Galactic plane scattering
screen, more particle acceleration leads to more optically thin jet
emission, which in turn increases the size of the photosphere at a
given frequency.  Thus M81*'s jets, with radiating particles
accelerated into a canonical power-law, would be predicted to
have a much larger photosphere at the same frequency than Sgr A*'s.  

The equipartition parameter, $k$, in M81* is very similar to the
values found in XRBs, favoring mild magnetic domination of the
internal energy.  Sgr A* on the other hand seems to be the only source
we have studied so far which favors a stronger magnetic energy density
($k\sim15$), bucking the trend seen here in M81* and in XRBs for a
correlation between total jet power and equipartition parameter.
Because we have no information about the non-thermal X-ray spectrum of
Sgr A* in quiescence, however, this parameter may not be well
constrained, but it is something to keep in mind for future
explorations.

The fitted jet nozzle scale height in M81* is not well-constrained by
these fits.  However, the fits that include upper limits as estimates
of the IR through optical spectra clearly select out a more elongated
base/corona.  Thus to strongly constrain the value of the nozzle scale
height, further simultaneous millimeter through IR/optical
observations are necessary.

Looking at the Maxwellian model fits, which have been applied to both
Sgr A* and hard state XRBs, we see the same trends when comparing to
the Sgr A* fits.  For the MXW cases, we can also compare acceleration
efficiency by the need for a power-law tail in the best-fit electron
spectrum.  Both M81* and XRBs are consistent with a high rate, here
frozen to 75\%, as compared to Sgr~A* where only a small fraction of
particles are accelerated.  In Sgr~A*, the contribution of the tail
also is minimized by acceleration occurring at quite large distances
from the jet base.  In contrast, the acceleration in M81* occurs at
the same location as seen in XRBs ($\sim 10-100\, r_g$).  In fact,
even though we are probing a fairly low fractional Eddington
luminosity compared to the previously fit XRBs, \textsl{nearly all
fitted parameters for the MXW fits to M81* fall into the exact same
ranges as those seen in hard state XRBs}.  The only major parameter
difference is the electron temperature, which is a factor of $\sim
2-3$ lower in XRBs.  These results thus provide very strong support for the
mass-scaling of accretion physics, at least for weakly accreting black
holes.

By comparing the fits to individual data sets, we can look for
meaningful correlations in the parameters on timescales impossible to
probe in XRBs, as weeks to months in M81* would correspond roughly to
sub-second variations in a typical XRB.  Overall there are more
obvious correlations among the MXW fit parameters than compared to the
PL fit parameters. This fact leads us to consider the MXW fits as more
likely probing more real physical effects.  However, simultaneous
constraints in the IR through UV are necessary to conclusively break
this degeneracy.  A selection of the strongest correlations are shown
in Figs.~\ref{eddratpars} \& \ref{equippars}.

Given the lack of simultaneous constraints on the IR through UV range,
it is somewhat surprising to see some trends linking intrinsic jet
parameters to the accretion disk parameters.  There is a clear
anti-correlation between the disk flux and the energy input to the
jet.  This may in fact be similar to the observed anti-correlation
between soft and hard X-ray fluxes seen in hard state XRBs, such as
Cyg~X-1 \citep[see][]{Wilmsetal2006}.  For M81*, the higher disk flux is
driven by an increase in the fitted disk temperature, but given that
we do not have simultaneous data directly in the spectral region where
the disk spectrum is most prominent (see Fig. 14), we cannot in fact
be sure that the correlation is not systematic.

More interesting are the correlations detected between the
equipartition parameter $k$ and other fitted parameters.  Along with
the total normalized power $N_{\rm jet}$, $k$ seems to be one of the
most important parameters for the jet model.  Because it dictates the
distribution of energy between the radiating particles and the
magnetic fields, it effects the synchrotron/inverse Compton ratio, and
can in some sense compensate for losses in the particle energy density
due to, for instance, temperature decreases.  As the temperature goes
down, more energy is needed is needed in the magnetic energy to
maintain the same synchrotron emission, thus requiring an increase in
$k$.  Similarly, a wider jet base results in a lower particle number
density and thus $k$ must again be increased to maintain the same
radiative fluxes.  The most interesting correlation, however, is that
between the jet power and the equipartition.  We seem to be seeing a
trend towards stronger magnetic powers relative to the radiating
particles with increasing total jet power.  With the exception of Sgr
A* (but see above for a reason to possibly discount this source in
this regard), this trend is also seen in individual fits to hard state
XRBs \citep[][; Maitra et al. in
prep.]{MarkoffNowakWilms2005,Migliarietal2007,Galloetal2007}.  One
possible interpretation of this is that the magnetic fields are more
efficiently generated at higher accretion rates.  After increasingly
building up, an explosive release of this energy at some critical
accretion rate may be responsible for driving the transient, and much
more relativistic, ejecta seen in transition to the XRB hard state.

\section{Conclusions}\label{sec:discuss}

Our simultaneous broadband campaign on the LLAGN M81* has generated
five individual spectra, spread over 6 months, as well as a combined
spectrum, that can be readily compared to other LLAGN, such as Sgr A*,
as well as hard-state, weakly accreting X-ray binaries.  These data
definitively confirm for the first time many species of line
emission from the accretion flow of an LLAGN \citep{Youngetal2007},
and we have confirmed previous detections of variability across the
entire spectrum.  The radio through submm in particular shows
significant levels of both intraday and longer term variability.  We
also see indications for adiabatically decaying ``flares'' moving out
along the jets.

The simultaneous, broadband nature of these data has allowed us to
fit the spectra with an outflow-dominated model developed for hard
state XRBs, that also has been used to understand our extremely
subluminous Galactic nucleus, Sgr A*.  We find several interesting
results based on these spectral fits.  Compared to Sgr A*, M81* is not
only much more luminous, but also more of its accretion energy is
funneled into accelerating and maintaining power-law distributions of
the radiating particles.  Otherwise, the geometry and particle
thermal/minimum temperatures seem to be very consistent between these
two sources.  We are currently conducting a campaign of similar scope
for the LLAGN NGC~4258 (Nowak et al., Reynolds et al., in prep.), which
will provide a third object to this ``sample'' of extensive,
simultaneous multi-wavelength datasets for LLAGN.

The most remarkable result of our modeling is the discovery that M81*
seems to behave just like a hard state XRB, despite it being over six
orders of magnitude more massive, and accreting at a fractional
Eddington luminosity somewhat lower than for the jet model fits to
XRBs.  The best fit parameters all fall into the same range as those
found for XRBs, with the exception of the particle initial
temperature/minimum energy, and the nozzle length.  We do not consider
the latter significant, however, since it cannot be well constrained
by this data set.  The temperature difference, on the other hand,
could be more notable since it is shared with fits from other LLAGN
\citep{Markoffetal2001,Yuanetal2002}.  If particles in LLAGN
enter the jets with a factor of 2--3 higher temperature than XRBs,
this could be due to the lower cooling rates for the comparatively
less compact LLAGN jets given the same power and size in mass-scaling
units of $L_{\rm Edd}$ and $r_g$.  Our results provide an
independent confirmation of the mass-scaling accretion physics
suggested by the fundamental plane of black hole accretion described
in \S~\ref{sec:model}.

Finally, given the $>20\%$ variations seen in the radio within the
campaign, as well as in comparison with prior observations, and the
factor of $\sim5$ times smaller average X-ray flux, it is clear that
the differences between average/non-simultaneous measurements and
simultaneous multiwavelength observations may be quite important for
LLAGN, and perhaps AGN as well.  In particular these variations
becomes relevant for the fundamental plane of black hole accretion.
Sub-Eddington accreting black holes show a correlation with slope
$\approx0.6$--0.7 between the logarithms of the radio and X-ray
luminosities, with an effective mass-dependent normalization.  The
exact values of the coefficients are important for placing stringent
limits on the processes responsible for the emission.  To date, the
coefficients primarily have been determined from samples of AGN/LLAGN
with non-simultaneous or averaged values for the radio and X-ray
luminosities.  Given the now confirmed significant broadband
variability of M81*, taking average values will likely lead to
incorrect determinations for the correlation coefficients.  To
quantify this possibility, we use the quasi-simultaneously measured
radio and X-ray luminosities from this campaign and redo the
fundamental plane statistical analysis of \cite{Markoff2005}. We find
a difference of 8\% in the radio/X-ray correlation slope, and a
difference of 50\% in the mass-dependence coefficient!  Our results
therefore strongly argue for some level of care in conclusions based
on non-(quasi)simultaneous observations.

\acknowledgements

We acknowledge the Guaranteed Time program of the {\em Chandra
  Observatory}, without which we would not have had an opportunity to
  pursue this campaign.  We also thank the staff of all participating
  observatories, who made these observations possible.  The GMRT is
  run by the National Center for Radio Astrophysics of the Tata
  Institute of Fundamental Research.  M.A.N. is supported by NASA
  Grant SV3-73016. PC gratefully acknowledges support from the
  National Radio Astronomy Observatory (NRAO) through a Jansky
  Fellowship.  NRAO is a facility of the National Science Foundation
  operated under cooperative agreement by Associated Universities,
  Inc..


\clearpage

\begin{deluxetable}{lccc}
\tablewidth{0pt}
\tablecaption{Log of 2005 February 23--25 Observations \label{febtimes}}
\tablehead{\colhead{Instrument} & {Energy/$\nu$} &
  \colhead{Start Time (UT)} & \colhead{End Time (UT)}}
\startdata
GMRT & 235/610 MHz & 23 Feb 14:15 & 23 Feb 22:57\\
GMRT & 1390 MHz & 24 Feb 13:56 & 24 Feb 16:23\\
VLA & 1.4/8.4/22/43 GHz & 24 Feb 05:00 & 24 Feb 12:30\\
PdBI & 115/230 GHz & 24 Feb 01:11 & 24 Feb 19:45\\
SMA & 345 GHz & 24 Feb 04:44 & 24 Feb 17:20\\
Lick & K/J/H band & 24 Feb 08:13 & 24 Feb 11:40\\
Lick & H band & 25 Feb 09:28 & 25 Feb 09:49\\
Chandra HETGS & 0.5--8 keV & 24 Feb 06:56 & 24 Feb 20:50 
\enddata
\tablecomments{In this and Tables~\ref{jultimes}--\ref{augtimes}, 
 the observing times
  represent the beginning and end of dedicated observational
  time. Integrated time is less than the full period, as calibration
  was performed intermittently during the intervals.  The {\em
  Chandra} ObsId for this observation was 6174.}
\end{deluxetable}

\begin{deluxetable}{lccc}
\tablewidth{0pt}
\tablecaption{Log of 2005 July 12--16 Observations \label{jultimes}}
\tablehead{\colhead{Instrument} & {Energy/$\nu$} &
  \colhead{Start Time (UT)} & \colhead{End Time (UT)}}
\startdata
GMRT & 235/610 MHz & 12 Jul 06:31 & 12 Jul 13:16\\
GMRT & 1390 MHz & 15 Jul 04:05 & 15 Jul 08:11\\
VLA & 1.4/8.4/22/43 GHz & 13 Jul 04:56 & 13 Jul 23:00\\
VLBA & 8.4 GHz & 13 Jul 17:30 & 14 Jul 05:30\\
PdBI & 80.5/241.4 GHz & 14 Jul 06:50 & 15 Jul 13:50\\
SMA & 230 GHz & 15 Jul 21:25 & 16 Jul 04:38\\
{\em Chandra HETGS} & 0.5--8 keV & 14 Jul 01:44 & 14 Jul 17:28\\
{\em Chandra HETGS} & 0.5--8 keV & 14 Jul 19:25 & 15 Jul 13:20
\enddata
\tablecomments{VLBA imaging at 22 GHz and SMA imaging at 230 GHz both
  failed due to poor weather. The {\em Chandra} ObsId for this observation were 6346 and 6347. }
\end{deluxetable}

\begin{deluxetable}{lccc}
\tablewidth{0pt}
\tablecaption{Log of 2005 July 18--20 Observations \label{jultimes2}}
\tablehead{\colhead{Instrument} & {Energy/$\nu$} &
  \colhead{Start Time (UT)} & \colhead{End Time (UT)}}
\startdata
VLA & 1.4/8.4/22/43 GHz & 19 Jul 16:44 & 19 Jul 23:07\\
PdBI & 86.2/218.2 GHz & 19 Jul 23:17 & 20 Jul 16:07\\
SMA & 230 GHz & 18 Jul 21:25 & 19 Jul 03:25\\
{\em Chandra HETGS} & 0.5--8 keV & 19 Jul 14:25 & 20 Jul 15:25
\enddata
\tablecomments{The {\em Chandra} ObsId for this observation was 5601.}
\end{deluxetable}

\begin{deluxetable}{lccc}
\tablewidth{0pt}
\tablecaption{Log of 2005 August 14--15 Observations \label{augtimes}}
\tablehead{\colhead{Instrument} & {Energy/$\nu$} &
  \colhead{Start Time (UT)} & \colhead{End Time (UT)}}
\startdata
VLA & 1.4/8.4/22/43 GHz & 14 Aug 13:19 & 14 Aug 19:15\\
SMA & 230 GHz & 14 Aug 19:37 & 15 Aug 01:30\\
{\em Chandra HETGS} & 0.5--8 keV & 14 Aug 09:50 & 14 Aug 20:57
\enddata
\tablecomments{PdBI experienced bad weather and no data were taken.  The {\em Chandra} ObsId for this observation was 5600.}
\end{deluxetable}

\begin{deluxetable}{lcccccc}
\tablewidth{0pt}
\tablecaption{Results of the GMRT Observations \label{m81_obs}}
\tablehead{\colhead{Date of}   & \colhead{Frequency} & 
\colhead{No. of good} & \colhead{\% of good} &  \colhead{Resoln.} & \colhead{Flux density} & \colhead{rms} \\
\colhead{Observn}  & \colhead{MHz}    &  
\colhead{Antennae} & \colhead{Data} & \colhead{$''$} & \colhead{mJy}     & \colhead{mJy}}
\startdata
2005 Feb 24 & 1390 &27&75  & 5x3 & $80.1\pm2.1$   & 0.24\\
2005 Jul 15 & 1390 &26 & 80 & 5x3   & $114.8\pm1.1$   & 0.32\\
\hline
2005 Feb 23 & 610 &27 &80  & 8x7  & $67.4\pm1.3$   & 0.45\\
2005 Jul 12 & 610 &29 & 90  & 8x6  & $76.3\pm1.2$   & 0.42\\
2005 Jul 26 & 610 &27& 75  & 9x5  & $72.4\pm3.1$   & 0.49\\
\hline
2005 Feb 23 & 235 &30 & 55  & 18x13 & $93.5\pm14.9$ & 4.95\\
2005 Jul 12 & 235  &28& 70 & 19x12 & $61.8\pm20.9$   & 3.80\\
\enddata
\end{deluxetable}

\begin{deluxetable}{lccc}
\tablewidth{0pt}
\tablecaption{Comparison of calibration observations of SN 1993J for
  GMRT and VLA \label{sn1993}}
\tablehead{\colhead{Date} & \colhead{Instrument} &  
\colhead{Frequency (GHz)} & \colhead{Flux density (mJy)}} 
\startdata
2005 Jul 14  & VLA & 1.4& $9.64\pm0.56$  \\
2005 Jul 14  & VLA & 8.4 & $2.43\pm0.29$  \\
2005 Jul 15  & GMRT & 1.4  &$9.00\pm0.52$  \\
2005 Aug 14  & VLA & 1.4 &  $10.23\pm0.80$\\
2005 Aug 14  & VLA & 8.4 & $1.87\pm0.20$ \\
\enddata
\end{deluxetable}

\begin{deluxetable}{lccc}
\tablewidth{0pt}
\tablecaption{Results of the VLA Observations\label{vla_obs}}
\tablehead{\colhead{Date of}   & \colhead{Frequency} & \colhead{Flux density} & \colhead{rms} \\

\colhead{Observn}  & \colhead{GHz}   & \colhead{mJy}     & \colhead{mJy}}
\startdata
2005 Feb 24 & 43   & --  & --\\
2005 Jul 13 & 43   & $66.5\pm8.5$  & \\
2005 Jul 19 & 43   & $143.4\pm3.1$  &\\
2005 Aug 14 & 43   & $82.1\pm2.4$  &\\
\hline
2005 Feb 24 & 22    &$101\pm20$   & \\
2005 Jul 13 & 22   & $86.4\pm5.8$  & \\
2005 Jul 19 & 22   & $133.6\pm0.6$  &\\
2005 Aug 14 & 22   & $109.6\pm1.4$  &\\
\hline
2005 Feb 24 & 8.4   &$176\pm20$   & \\
2005 Jul 13 & 8.4   &$89.7\pm2.6$   & \\
2005 Jul 19 & 8.4   &$105.1\pm0.4$   &\\
2005 Aug 14 & 8.4   &$112.5\pm0.1$   &\\
\hline
2005 Feb 24 & 1.4   & $75\pm10$  & \\
2005 Jul 13 & 1.4   &$91.9\pm0.3$   & \\
2005 Jul 19 & 1.4   &$92.6\pm0.9$   &\\
2005 Aug 14 & 1.4   &$89.5\pm0.6$   &\\
\enddata
\tablecomments{Due to bad weather, no data were taken at 43 GHz in
  February.}
\end{deluxetable}

\begin{deluxetable}{lccccc}
\tablewidth{0pt}
\tablecaption{Results of the PdBI observations\label{tab:pdbi}}
\tablehead{\colhead{Date of}   & \colhead{Frequency} & 
\colhead{No. of good} & \colhead{\% of good} & \colhead{Flux density} & \colhead{rms} \\

\colhead{Observn} & \colhead{GHz} & \colhead{Antennae} &
\colhead{Data} & \colhead{mJy} & \colhead{mJy}} 
\startdata 
2005 Feb 24 & 115.3 & 6 & 90\% & $88.0\pm11.7$ & --\\ 
2005 Jul 14 & 80.5 &5 & 50\% & $241.2\pm33.8$ & \\ 
2005 Jul 19 & 86.2 &5 &98\% & $118.7\pm11.4$ &\\ 
\hline 
2005 Feb 24 & 230.5 & 6 & 90\%&$85.6\pm17.8$ & \\ 
2005 Jul 14 & 241.4 &5 & 30\%& $181.2\pm39.1$ & \\ 
2005 Jul 19 & 218.2 &5 & 90\% & $74.8\pm13.3$ &\\
\enddata \tablecomments{The listed flux densities result from the mean
and standard deviation of the lightcurves obtained during the
observations \citep{Schoedeletal2007}.}
\end{deluxetable}

\begin{deluxetable}{lccc}
\tablewidth{0pt}
\tablecaption{Results of the SMA Observations\label{sma_obs}}
\tablehead{\colhead{Date of}   & \colhead{Frequency} & 
\colhead{No. of } & \colhead{Flux density} \\

\colhead{Observn}  & \colhead{GHz}    &  
\colhead{Antennae} & \colhead{mJy} }
\startdata
2005 Feb 24 & 340.7  & 7 & $378.7\pm70.0$  \\
2005 Jul 18 & 225.4  & 7 & $182.8\pm36.0$  \\
2005 Aug 14 & 225.4  & 7 & $91.5\pm15.3$   \\
\hline
\enddata
\end{deluxetable}

\begin{deluxetable}{ll}
\tablecaption{Summary of jet model parameters \label{tab:params}}
\tablecolumns{2}
\tablehead{\colhead{Parameter} & \colhead{Meaning}}
\startdata
\multicolumn{2}{c}{\sc Both models} \\
$N_{\rm jet}$ & Jet normalized power  \\
$r_0$ & Size of jet base \\
$T_e$ & Temperature of particles entering base of jet \\
$p$ &  Energy index of electron power-law tail ($N_e\propto E_e^{-p}$)\\
$k$ & Equipartition parameter: the ratio of energy density in the magnetic field \\
&  to energy density in the radiating particles \\
$h_{\rm ratio}$ & Ratio of jet nozzle length to nozzle radius \\
$L_{\rm d}$ & Accretion disk bolometric luminosity \\
$T_d$ & Innermost accretion disk temperature \\
 \\
\multicolumn{2}{c}{\sc Powerlaw (PL) model} \\
$\gamma_{\rm max}$ & Maximum Lorentz factor of electron distribution\\
 \\
\multicolumn{2}{c}{\sc Maxwellian (MXW) model} \\
$z_{\rm acc}$ & Location at which particles are first accelerated to produce powerlaw tail\\
$\epsilon_{\rm sc}$ & Particle acceleration efficiency parameter; $\epsilon_{\rm sc} \equiv (u_{\rm sh}/c)^2 / \xi$ where $u_{\rm sh}$ is the relative \\
& velocity of the shock to the plasma in the shock frame, and $\xi$ is the \\
& ratio between the mean free path to the gyroradius.
\enddata
\end{deluxetable}

\begin{deluxetable}{cccccc} 
\setlength{\tabcolsep}{0.03in} 
\tabletypesize{\scriptsize} 
\tablewidth{0pt} 
\tablecaption{Jet Model Fits to M81$^*$ (PL) \label{tab:PL_FINAL_a}} 
\tablehead{ \colhead{Obs ID}           
           & \colhead{L$_{\rm jet}$}   
           & \colhead{$r_{0}$}          
           & \colhead{$T_e$}            
           & \colhead{$p$}              
           & \colhead{$k$}              
           \\                          
           & ($10^{-5}$ L$_{\rm Edd}$)  
           & ($GM/c^2$)                  
           & ($10^{10}$\,K)             
           }                             
\startdata  
6174 
           & \errtwo{   4.207}{   0.013}{   0.019} 
           & \errtwo{   4.536}{   0.028}{   0.000} 
           & \errtwo{   10.00}{    0.02}{    0.03} 
           & \errtwo{   2.729}{   0.003}{   0.003} 
           & \errtwo{   1.019}{   0.010}{   0.015} 
           \\  
\tabspace 
6346 
           & \errtwo{   4.318}{   0.020}{   0.019} 
           & \errtwo{   4.655}{   0.016}{   0.075} 
           & \errtwo{   10.07}{    0.03}{    0.03} 
           & \errtwo{   2.797}{   0.003}{   0.171} 
           & \errtwo{   1.222}{   0.018}{   0.017} 
           \\  
\tabspace 
6347 
           & \errtwo{   4.316}{   0.019}{   0.016} 
           & \errtwo{   4.657}{   0.013}{   0.078} 
           & \errtwo{   10.06}{    0.03}{    0.03} 
           & \errtwo{   2.798}{   0.002}{   0.005} 
           & \errtwo{   1.217}{   0.017}{   0.016} 
           \\  
\tabspace 
5601 
           & \errtwo{   4.295}{   0.016}{   0.015} 
           & \errtwo{   5.190}{   0.015}{   0.002} 
           & \errtwo{    9.90}{    0.02}{    0.02} 
           & \errtwo{   2.798}{   0.002}{   0.202} 
           & \errtwo{   1.471}{   0.017}{   0.016} 
           \\  
\tabspace 
5600 
           & \errtwo{   4.357}{   0.027}{   0.026} 
           & \errtwo{   4.936}{   0.019}{   0.001} 
           & \errtwo{    9.89}{    0.04}{    0.04} 
           & \errtwo{   2.628}{   0.007}{   0.006} 
           & \errtwo{   1.075}{   0.021}{   0.020} 
           \\  
\tabspace 
\tabspace 
\tableline 
\tabspace 
\tabspace 
ALL\tablenotemark{a} 
           & \errtwo{   4.238}{ 0.009}{ 0.004} 
           & \errtwo{   4.553}{  0.003}{  0.000} 
           & \errtwo{   10.55}{ 0.01}{ 0.01} 
           & \errtwo{   2.800}{0.028}{0.002} 
           & \errtwo{   1.028}{ 0.005}{ 0.004} 
           \\  
\tabspace 
\tableline 
\tabspace 
\tabspace 
ALL\tablenotemark{b} 
           & \errtwo{   4.211}{ 0.012(+0.146)}{ 0.007(-0.003)} 
           & \errtwo{   4.553}{  0.005(+0.637)}{  0.009(-0.017)} 
           & \errtwo{   10.56}{  0.02(-0.48)}{  0.01(-0.66)} 
           & \errtwo{   2.800}{0.000(-0.001)}{0.003(-0.172)} 
           & \errtwo{   1.029}{ 0.009(+0.442)}{ 0.004(-0.009)} 
           \\  
\enddata 
\tablecomments{} 
\tablenotetext{a}{HEG data from each ObsID combined, 
and MEG data from each ObsID combined.  The radio data were left 
uncombined, and upper limits from non-simultaneous IRO/UV observations
           were entered as detections with 20\% error bars (see text). } 
\tablenotetext{b}{HEG data from each ObsID combined, 
and MEG data from each ObsID combined.  The radio data were left 
uncombined, and only the non-simultaneous \textsl{HST} data were
  included with 20\% error bars.  Error bars in parentheses show the differences with
respect to the minimum and maximum from the fits to individual ObsIds. } 
\end{deluxetable}

\begin{deluxetable}{ccccccc} 
\setlength{\tabcolsep}{0.03in} 
\tabletypesize{\scriptsize} 
\tablewidth{0pt} 
\tablecaption{Jet Model Fits to M81$^*$ (PL), continued \label{tab:PL_FINAL_b}} 
\tablehead{ \colhead{Obs ID}          
           & \colhead{$h_{ratio}$}     
           & \colhead{$\Gamma$}       
           & \colhead{L$_{\rm d}$}    
           & \colhead{$T_{\rm d}$}    
           & \colhead{$\chi^2$/DoF}   
           & \colhead{$\chi^2_{\rm red}$}
           \\                         
           & &                          
           & ($10^{-7}$ L$_{\rm Edd}$) 
           & ($10^5$\,K)               
          }                             
\startdata  
6174 
           & \errtwo{  10.702}{   1.625}{   0.018} 
           & \errtwo{   34.68}{    1.19}{    0.46} 
           & \errtwo{     7.9}{     1.9}{     1.9} 
           & \errtwo{    0.48}{    0.18}{    0.10} 
           &      504.3/472 
           &     1.07 
           \\  
\tabspace 
6346 
           & \errtwo{  12.949}{   0.021}{   0.032} 
           & \errtwo{   26.46}{    1.17}{    0.97} 
           & \errtwo{    13.0}{     3.4}{     3.4} 
           & \errtwo{    0.50}{    0.19}{    0.11} 
           &      529.4/536 
           &     0.99 
           \\  
\tabspace 
6347 
           & \errtwo{  12.945}{   0.020}{   0.030} 
           & \errtwo{   27.92}{    1.29}{    0.96} 
           & \errtwo{    12.9}{     3.7}{     3.4} 
           & \errtwo{    0.53}{    0.19}{    0.12} 
           &      627.2/605 
           &     1.04 
           \\  
\tabspace 
5601 
           & \errtwo{  11.782}{   1.961}{   0.006} 
           & \errtwo{   30.25}{    1.24}{    1.00} 
           & \errtwo{    26.5}{     8.1}{     8.0} 
           & \errtwo{    1.11}{    0.39}{    0.21} 
           &      818.4/782 
           &     1.05 
           \\  
\tabspace 
5600 
           & \errtwo{   9.836}{   1.942}{   0.098} 
           & \errtwo{   21.12}{    0.75}{    0.74} 
           & \errtwo{    13.9}{     3.3}{     3.3} 
           & \errtwo{    0.49}{    0.16}{    0.11} 
           &      380.4/321 
           &     1.19 
           \\  
\tabspace 
\tabspace 
\tableline 
\tabspace 
\tabspace 
ALL\tablenotemark{a} 
           & \errtwo{  10.663}{1.241}{0.000} 
           & \errtwo{   32.34}{0.14}{1.39} 
           & \errtwo{    10.6}{ 1.8}{ 1.8} 
           & \errtwo{    0.60}{0.13}{0.07} 
           &     574.0/292  
           &    1.97  
           \\  
\tabspace 
\tableline 
\tabspace 
\tabspace 
ALL\tablenotemark{b} 
           & \errtwo{  11.412}{1.168(+1.538)}{0.749(-1.575)} 
           & \errtwo{   35.07}{ 0.58(-0.40)}{1.05(-13.95)} 
           & \errtwo{     7.4}{ 1.9(+19.1)}{ 1.6(+0.6)} 
           & \errtwo{    0.59}{ 0.17(+ 0.52)}{ 0.11(- 0.11)} 
           &     422.7/282  
           &    1.50  
           \\  
\enddata 
\tablecomments{} 
\tablenotetext{a}{HEG data from each ObsID combined, 
and MEG data from each ObsID combined.  The radio data were left 
uncombined, and upper limits from non-simultaneous IRO/UV observations
           were entered as detections with 20\% error bars (see text). } 
\tablenotetext{b}{HEG data from each ObsID combined, 
and MEG data from each ObsID combined.  The radio data were left 
uncombined, and only the non-simultaneous \textsl{HST} data were
  included with 20\% error bars.  Error bars in parentheses show the differences with
respect to the minimum and maximum from the fits to individual ObsIds. } 
\end{deluxetable}

\begin{deluxetable}{ccccccc} 
\setlength{\tabcolsep}{0.03in} 
\tabletypesize{\scriptsize} 
\tablewidth{0pt} 
\tablecaption{Jet Model Fits to M81$^*$ (MXSW) \label{tab:MXSW_FINAL_a}} 
\tablehead{ \colhead{Obs ID}           
           & \colhead{L$_{\rm jet}$}   
           & \colhead{$r_{0}$}          
           & \colhead{$T_e$}            
           & \colhead{$p$}              
           & \colhead{$k$}              
           & \colhead{$z_{acc}$}        
           \\                          
           & ($10^{-5}$ L$_{\rm Edd}$)  
           & ($GM/c^2$)                  
           & ($10^{10}$\,K)             
           & & & ($GM/c^2$)              
          }                              
\startdata  
6174 
           & \errtwo{   4.543}{   0.052}{   0.164} 
           & \errtwo{    4.63}{    0.52}{    1.61} 
           & \errtwo{    8.45}{    0.13}{    0.51} 
           & \errtwo{   2.424}{   0.024}{   0.002} 
           & \errtwo{   1.842}{   0.061}{   0.115} 
           & \errtwo{   142.0}{    14.0}{    19.3} 
           \\  
\tabspace 
6346 
           & \errtwo{   3.630}{   0.025}{   0.114} 
           & \errtwo{    1.78}{    0.25}{    0.24} 
           & \errtwo{   11.49}{    0.11}{    0.26} 
           & \errtwo{   2.407}{   0.012}{   0.002} 
           & \errtwo{   1.304}{   0.022}{   0.338} 
           & \errtwo{   131.6}{    33.4}{     9.6} 
           \\  
\tabspace 
6347 
           & \errtwo{   3.547}{   0.022}{   0.102} 
           & \errtwo{    1.96}{    0.26}{    0.04} 
           & \errtwo{   11.47}{    0.07}{    0.08} 
           & \errtwo{   2.413}{   0.002}{   0.001} 
           & \errtwo{   1.298}{   0.019}{   0.073} 
           & \errtwo{   124.4}{     6.3}{     2.5} 
           \\  
\tabspace 
5601 
           & \errtwo{   3.967}{   0.021}{   0.032} 
           & \errtwo{    3.13}{    0.07}{    0.09} 
           & \errtwo{   10.14}{    0.06}{    0.07} 
           & \errtwo{   2.401}{   0.001}{   0.001} 
           & \errtwo{   1.503}{   0.019}{   0.038} 
           & \errtwo{   192.2}{     0.8}{    38.7} 
           \\  
\tabspace 
5600 
           & \errtwo{   4.033}{   0.039}{   0.036} 
           & \errtwo{    2.88}{    0.14}{    0.38} 
           & \errtwo{   10.13}{    0.09}{    0.10} 
           & \errtwo{   2.403}{   0.002}{   0.002} 
           & \errtwo{   1.490}{   0.032}{   0.027} 
           & \errtwo{   193.5}{    49.8}{     0.3} 
           \\  
\tabspace 
\tabspace 
\tableline 
\tabspace 
\tabspace 
ALL\tablenotemark{a} 
           & \errtwo{   3.512}{0.007}{0.009} 
           & \errtwo{    2.42}{ 0.04}{ 0.05} 
           & \errtwo{   11.60}{ 0.02}{ 0.02} 
           & \errtwo{   2.412}{0.002}{0.001} 
           & \errtwo{   1.376}{0.006}{0.502} 
           & \errtwo{   143.9}{  0.1}{ 22.0} 
           \\  
\tabspace 
\tableline 
\tabspace 
\tabspace 
ALL\tablenotemark{b} 
           & \errtwo{   3.565}{ 0.011(+0.979)}{ 0.015(-0.018)} 
           & \errtwo{    2.40}{  0.06(+2.23)}{  0.08(-0.61)} 
           & \errtwo{   11.41}{  0.04(+0.08)}{  0.07(-2.96)} 
           & \errtwo{   2.409}{0.001(+0.015)}{0.001(-0.008)} 
           & \errtwo{   1.340}{ 0.010(+0.502)}{ 0.026(-0.042)} 
           & \errtwo{   144.4}{   1.5(+49.1)}{  22.5(-20.0)} 
           \\  
\enddata 
\tablecomments{} 
\end{deluxetable} 
\tablenotetext{a}{HEG data from each ObsID combined, 
and MEG data from each ObsID combined.  The radio data were left 
uncombined, and upper limits from non-simultaneous IRO/UV observations
           were entered as detections with 20\% error bars (see text). } 
\tablenotetext{b}{HEG data from each ObsID combined, 
and MEG data from each ObsID combined.  The radio data were left 
uncombined, and only the non-simultaneous \textsl{HST} data were
  included with 20\% error bars.  Error bars in parentheses show the differences with
respect to the minimum and maximum from the fits to individual ObsIds. }  

\begin{deluxetable}{cccccccc} 
\setlength{\tabcolsep}{0.03in} 
\tabletypesize{\scriptsize} 
\tablewidth{0pt} 
\tablecaption{Jet Model Fits to M81$^*$ (MXSW), continued \label{tab:MXSW_FINAL_b}} 
\tablehead{ \colhead{Obs ID}           
           & \colhead{$h_{ratio}$}      
           & \colhead{$f_{sc}$}         
           & \colhead{L$_{\rm d}$}     
           & \colhead{$T_{\rm d}$}     
           & \colhead{$\chi^2$/DoF}    
           & \colhead{$\chi^2_{\rm red}$}
           \\                          
           & &                           
           & ($10^{-7}$ L$_{\rm Edd}$)  
           & ($10^5$\,K)                
          }                              
\startdata  
6174 
           & \errtwo{  10.490}{   0.622}{   3.861} 
           & \errtwo{   245.8}{    47.9}{    21.3} 
           & \errtwo{     8.7}{     2.6}{     2.9} 
           & \errtwo{    0.35}{    0.16}{    0.22} 
           &      497.0/471 
           &     1.06     
           \\  
\tabspace 
6346 
           & \errtwo{   5.224}{   0.148}{   0.896} 
           & \errtwo{   516.2}{    59.8}{    43.8} 
           & \errtwo{    18.1}{     5.1}{     5.1} 
           & \errtwo{    0.76}{    0.21}{    0.04} 
           &      533.8/535 
           &     1.00     
           \\  
\tabspace 
6347 
           & \errtwo{   7.432}{   0.279}{   1.152} 
           & \errtwo{    63.1}{     7.7}{     4.8} 
           & \errtwo{    20.0}{     5.6}{     5.6} 
           & \errtwo{    0.82}{    0.27}{    0.15} 
           &      608.4/604 
           &     1.01     
           \\  
\tabspace 
5601 
           & \errtwo{  15.090}{   0.278}{   0.860} 
           & \errtwo{   157.8}{    15.7}{    12.9} 
           & \errtwo{    11.5}{     3.5}{     3.3} 
           & \errtwo{    0.51}{    0.21}{    0.13} 
           &      826.0/781 
           &     1.06     
           \\  
\tabspace 
5600 
           & \errtwo{  11.436}{   1.757}{   0.813} 
           & \errtwo{    92.8}{    17.9}{    15.7} 
           & \errtwo{    11.8}{     3.3}{     3.3} 
           & \errtwo{    0.49}{    0.20}{    0.12} 
           &      322.8/320 
           &     1.01     
           \\  
\tabspace 
\tabspace 
\tableline 
\tabspace 
\tabspace 
ALL\tablenotemark{a} 
           & \errtwo{   7.799}{0.063}{0.020} 
           & \errtwo{   206.1}{  1.1}{  0.1} 
           & \errtwo{    19.8}{ 1.6}{ 1.6} 
           & \errtwo{    0.82}{ 0.06}{ 0.06} 
           &     576.1/291  
           &    1.98      
           \\  
\tabspace 
\tableline 
\tabspace 
\tabspace 
ALL\tablenotemark{b} 
           & \errtwo{   7.586}{0.179(+7.504)}{0.191(-2.361)} 
           & \errtwo{   206.2}{  0.6(+310.0)}{  0.8(-143.1)} 
           & \errtwo{    19.9}{ 2.7(+ 0.1)}{  2.9(-11.3)} 
           & \errtwo{    0.83}{ 0.12(-0.01)}{ 0.09(- 0.48)} 
           &     379.4/281  
           &    1.35      
           \\  
\enddata 
\tablecomments{} 
\tablenotetext{a}{HEG data from each ObsID combined,
and MEG data from each ObsID combined.  The radio data were left 
uncombined, and upper limits from non-simultaneous IRO/UV observations
           were entered as detections with 20\% error bars (see text).} 
\tablenotetext{b}{HEG data from each ObsID combined, 
and MEG data from each ObsID combined. The radio data were left 
uncombined, and only the non-simultaneous \textsl{HST} data were
  included with 20\% error bars.  Error bars in parentheses show the differences with
respect to the minimum and maximum from the fits to individual ObsIds.  } 
\end{deluxetable}

\clearpage

\begin{figure*}
\includegraphics[angle=-90,width=\textwidth]{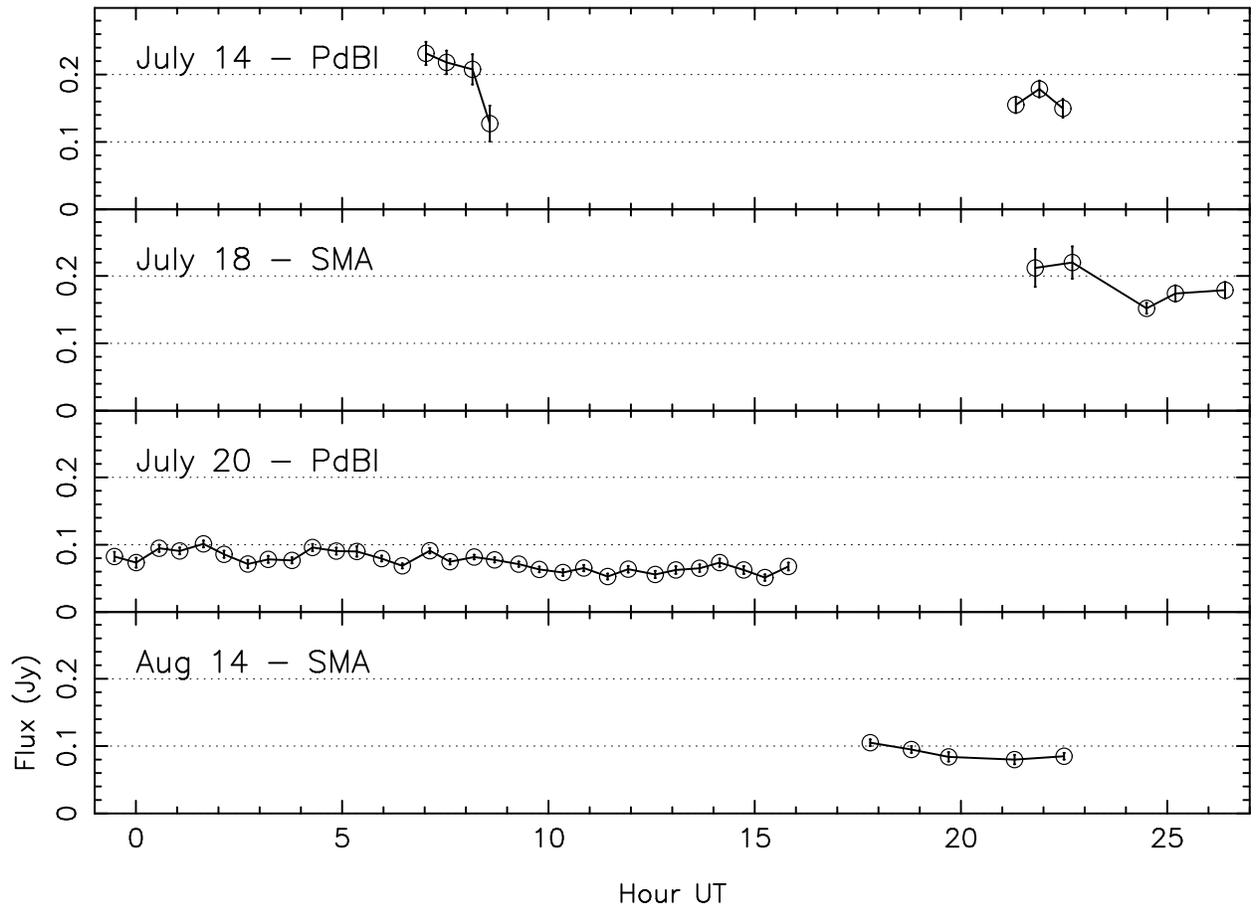}
\caption{Comparison of 1mm PdBI and SMA observations in July and August.   \label{1mmpdbsma}}
\end{figure*}  

\begin{figure*}
\epsscale{1}
\plotone{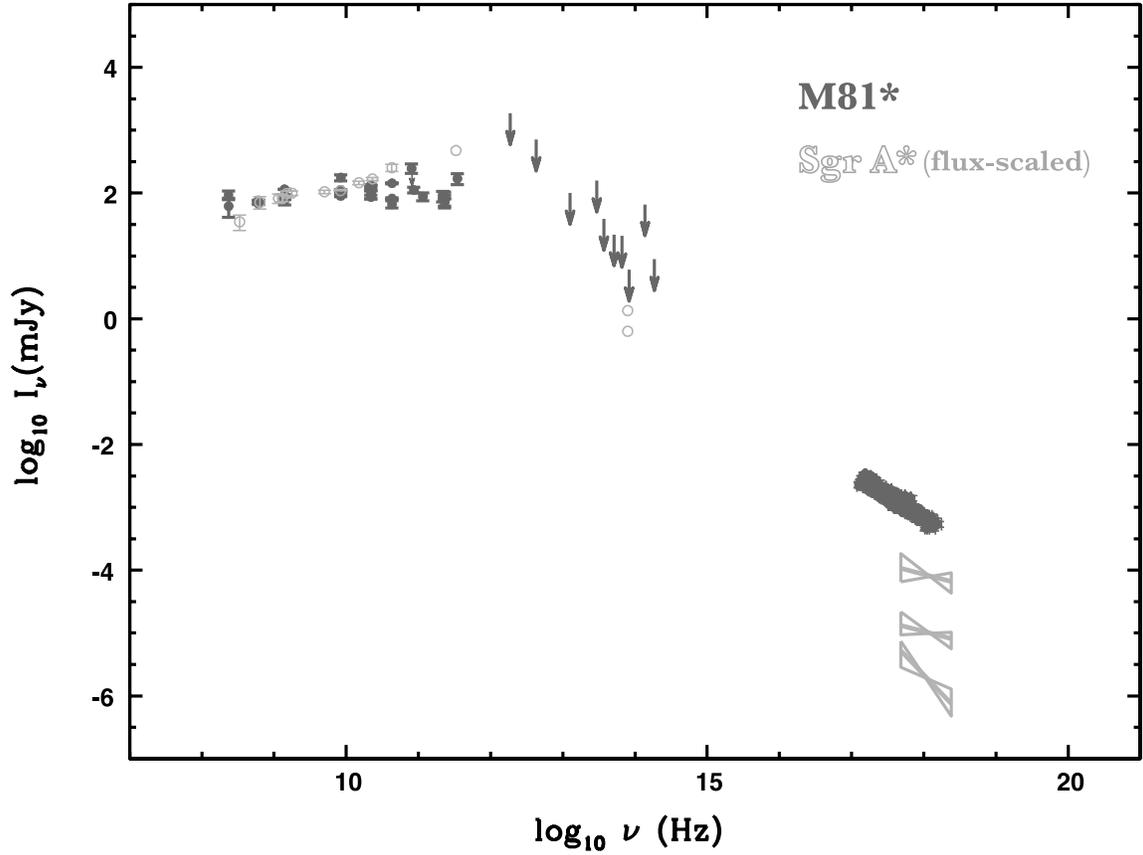}
\caption{Comparison between total combined broadband spectrum of M81*
  (as in Fig.~\ref{mongo_ul}) and Sgr A*, where the data for
  Sgr A* have been scaled downwards by roughly an order of magnitude
  so the VLA-band fluxes are the same.  The radio through NIR data for
  Sgr A* are the result of a simultaneous campaign presented in
  \cite{Anetal2005}.  The X-ray ``bowties'' represent the quiescent,
  average daily {\em Chandra} flare and highest {\em Chandra} flare
  detected, respectively
  \cite{Baganoffetal2001,Baganoff2003,Baganoffetal2003}.\label{m81sgra}}
\end{figure*}

\begin{figure*}
\epsscale{1}
\plotone{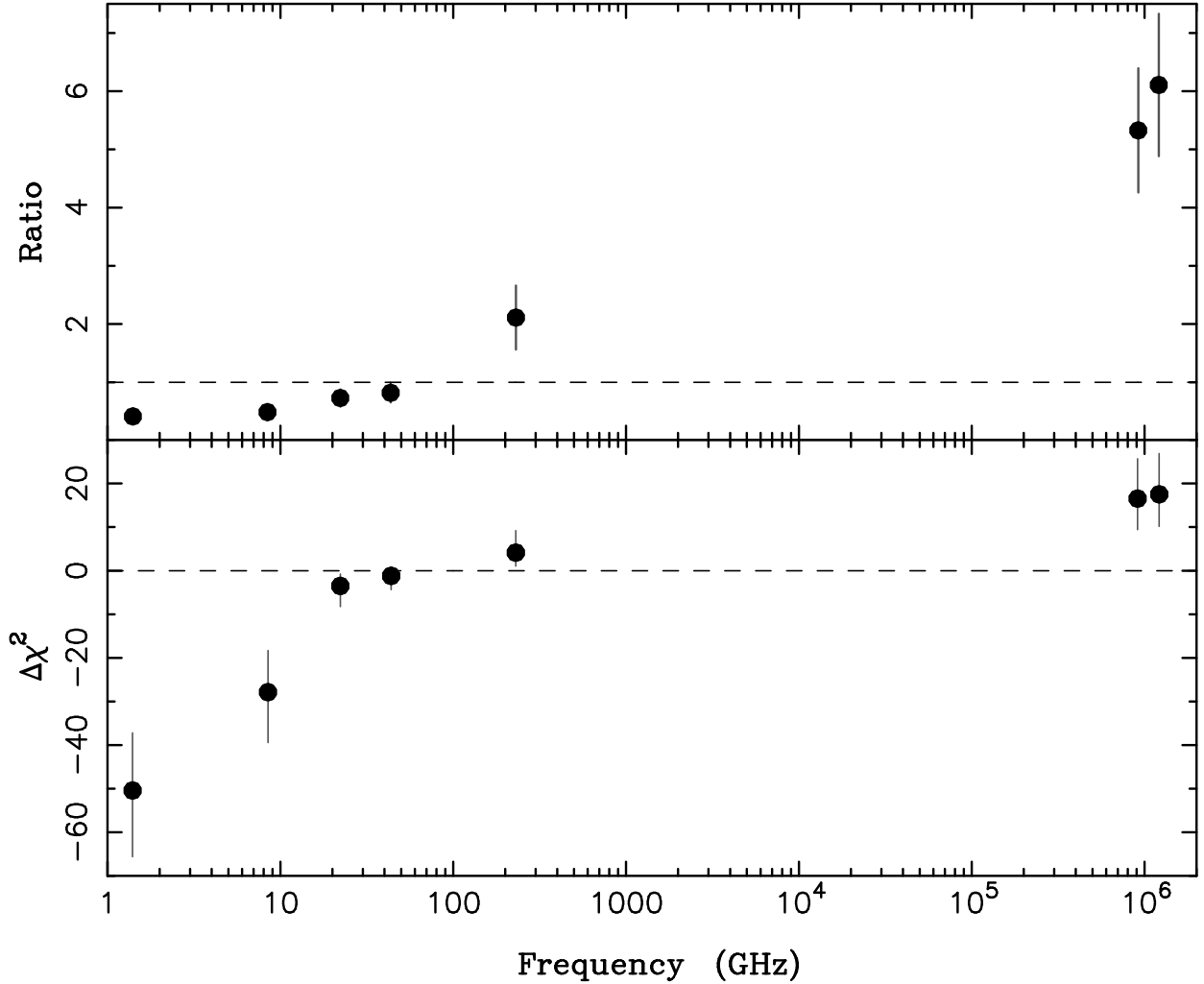}
\caption{As an illustration of the importance of broadband data for
  constraining jet models, we show here the radio data residuals for a
  fit to the X-ray data alone for Chandra ObsID 5600 (14 Aug).  Top
  panel: ratio of data to model.  Bottom panel: Change in $\chi^2$
  compared to the best fit presented below, where 20\% systematic
  error bars again have been added in quadrature to the radio data
  error bars.  This fit results in jet parameters $z_{acc} \approx
  1100$ and $h_{ratio} \approx 1.7$, ${\rm L_{jet}}$ and $r_{\rm 0}$
  are at twice the typical values found in the fits presented in the
  tables, and $T_e$ is at one third the typical value presented in the
  tables.  This fit is almost indistinguishable from the tabulated
  fits in the X-ray regime ($\Delta \chi^2=1.2$), but fails completely
  in the optical and radio regime. \label{radioresids}}
\end{figure*}

\begin{figure*}
\epsscale{1}
\plottwo{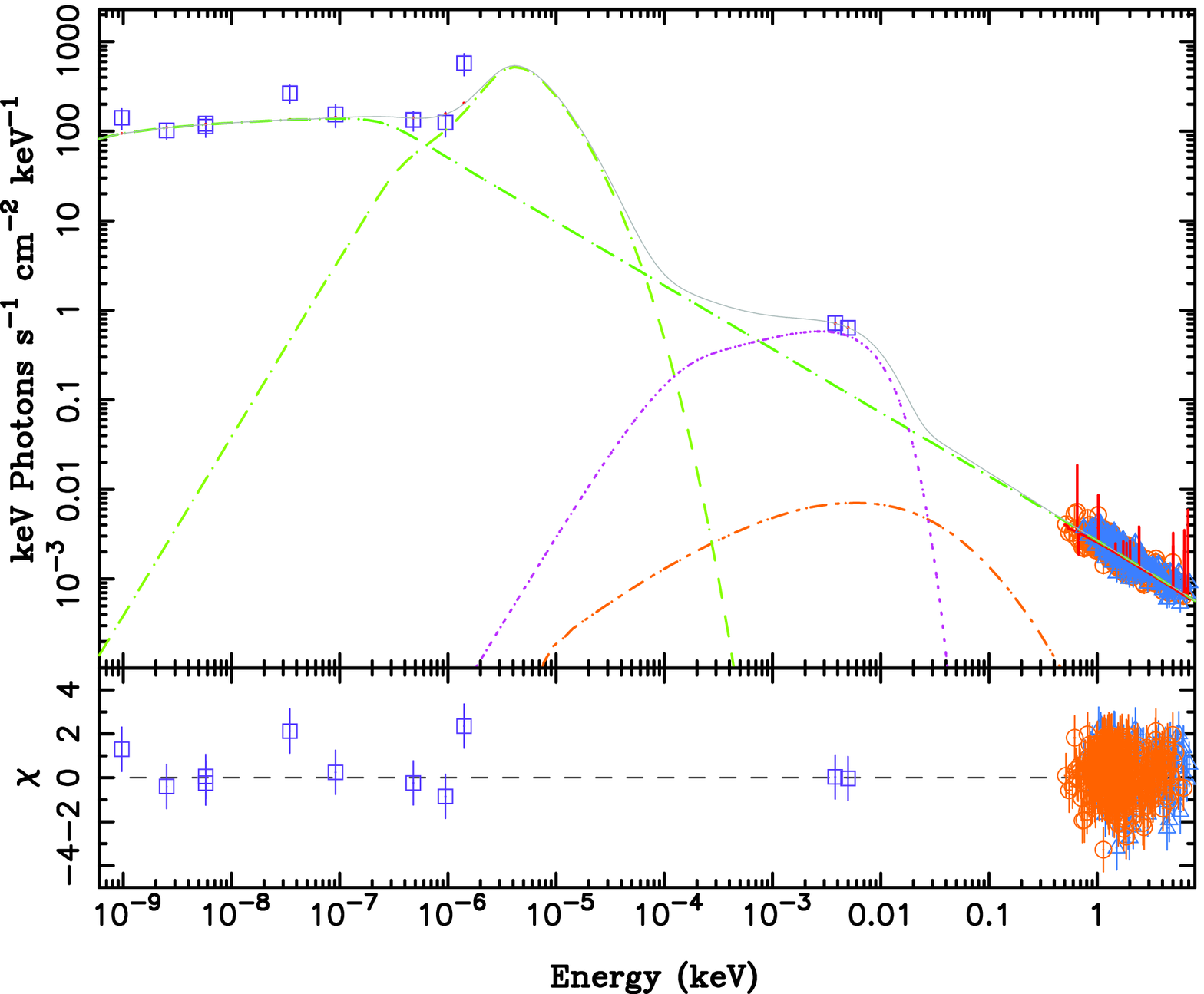}{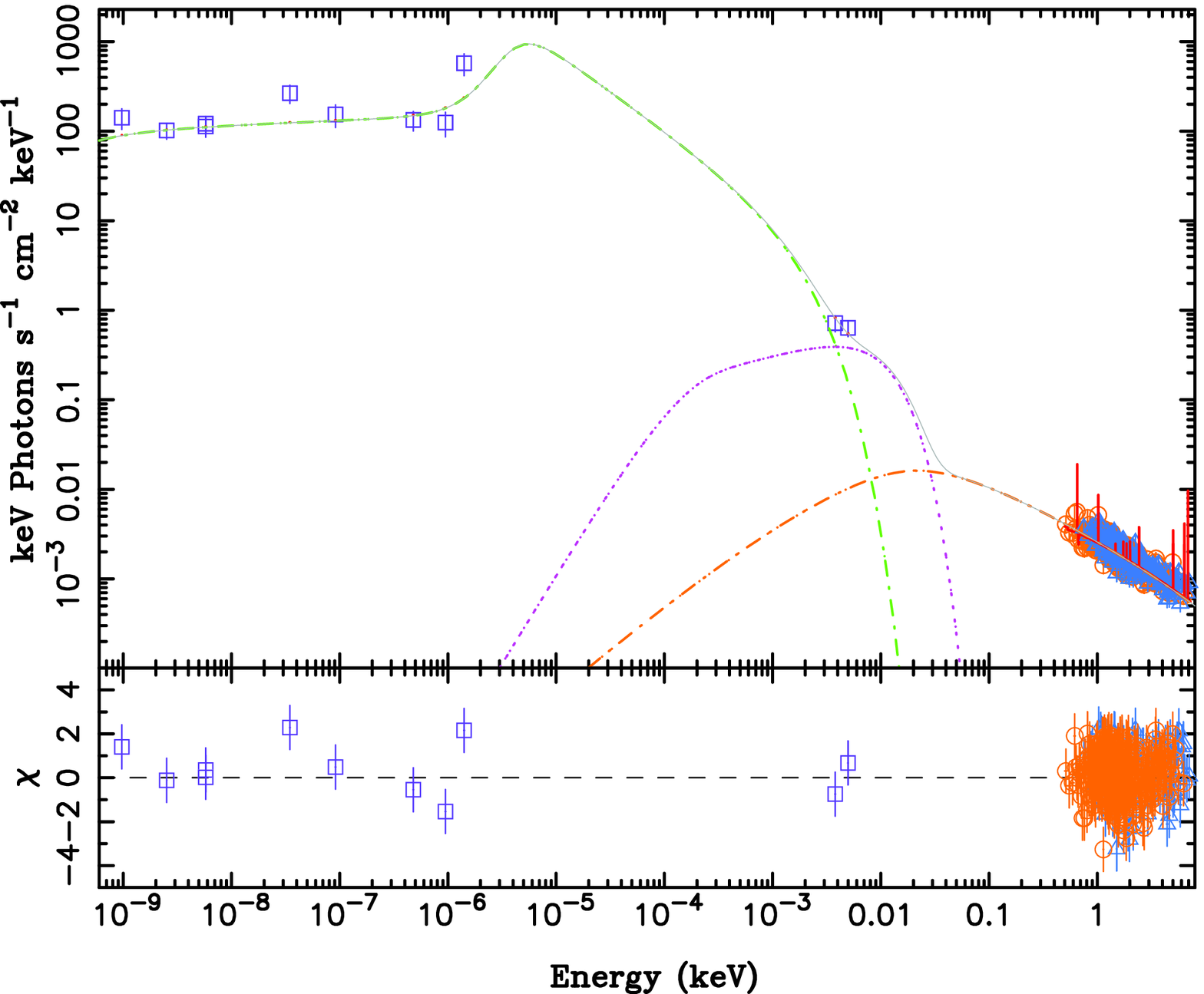}
\caption{Model fit to broadband February data with residuals. {\em
  Left:} Fit with initially Maxwellian leptonic distribution, {\em Right:} 
  leptonic power-law distribution fit.  The symbols represent the data,
  while the solid red line is the model fit in detector space.  The
  other components are not convolved with the detector matrices and
  serve only to indicate the contributing components to the continuum
  model.  Solid (grey): total spectrum, Dot-longdashed (light green):
  pre-acceleration inner jet synchrotron emission, Dot-longdashed
  (darker green): post-acceleration outer jet synchrotron,
  Dot-longdash-shortdash (orange): Compton emission from the inner jet
  (including external disk photons as well as synchrotron
  self-Compton), Dots-shortdash (magenta): thermal multicolor-blackbody
  disk model. \label{febfit} }
\end{figure*} 

\begin{figure*}
\epsscale{1}
\plottwo{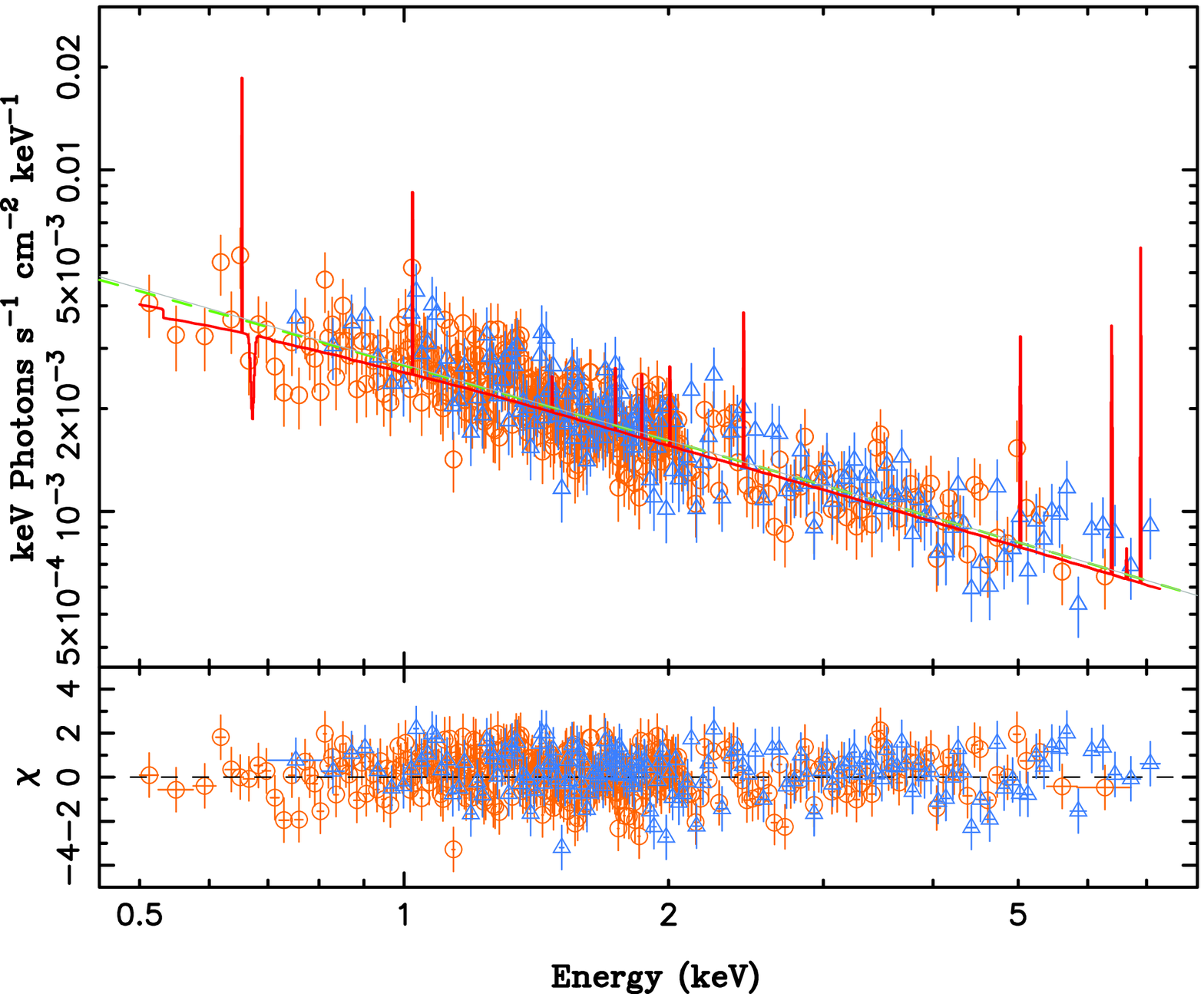}{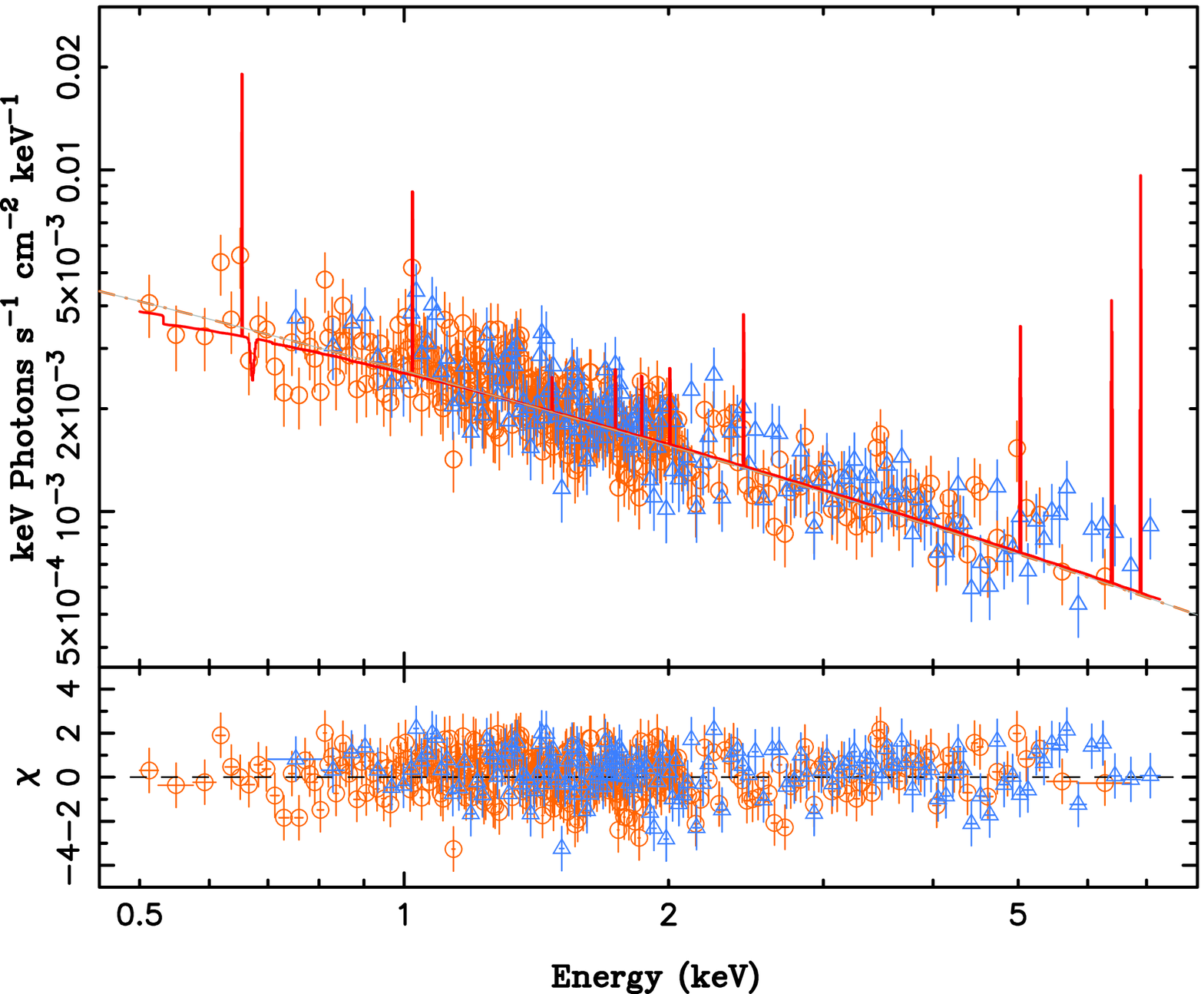}
\caption{Same as Fig.~\ref{febfit}, closeup view of X-ray band. An
  explanation of the included lines can be found in \cite{Youngetal2007}. \label{febfit_xray}}
\end{figure*} 

\clearpage

\begin{figure*}
\epsscale{1}
\plottwo{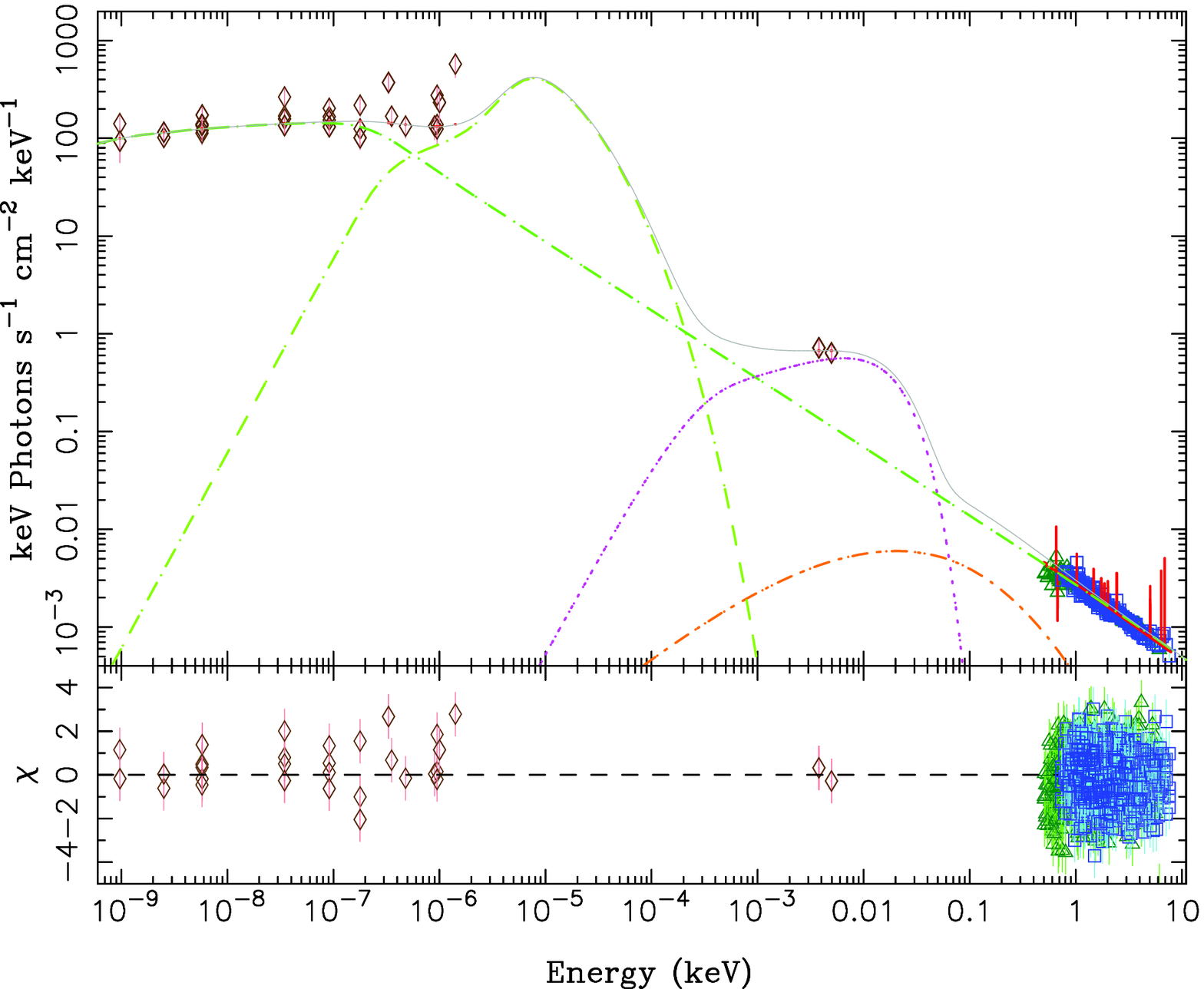}{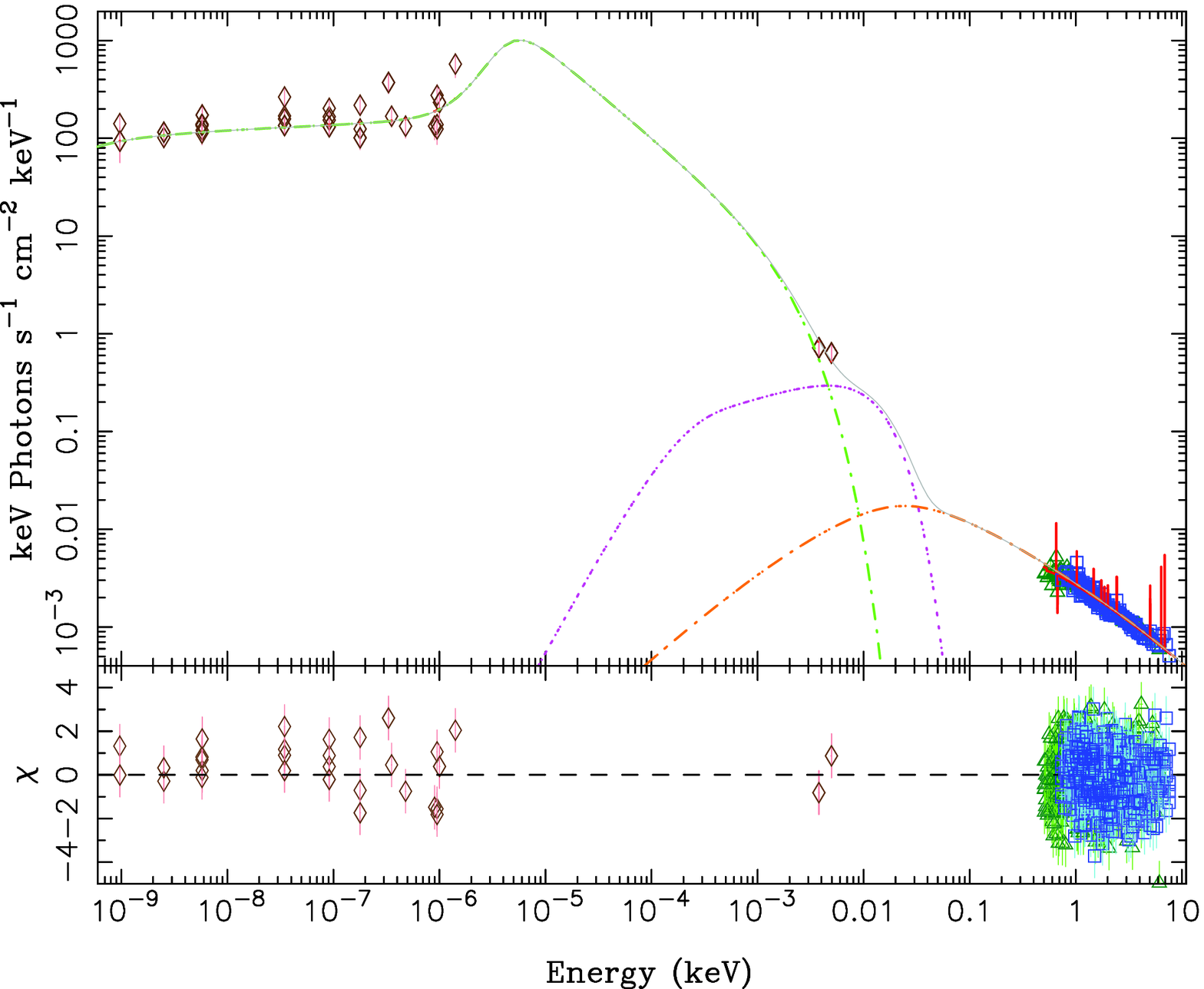}
\caption{Model fit to combined set of all broadband observations, with
  the same symbol/line definitions as Fig.~\ref{febfit}.\label{mongo}}
\end{figure*}

\begin{figure*}
\epsscale{1}
\plottwo{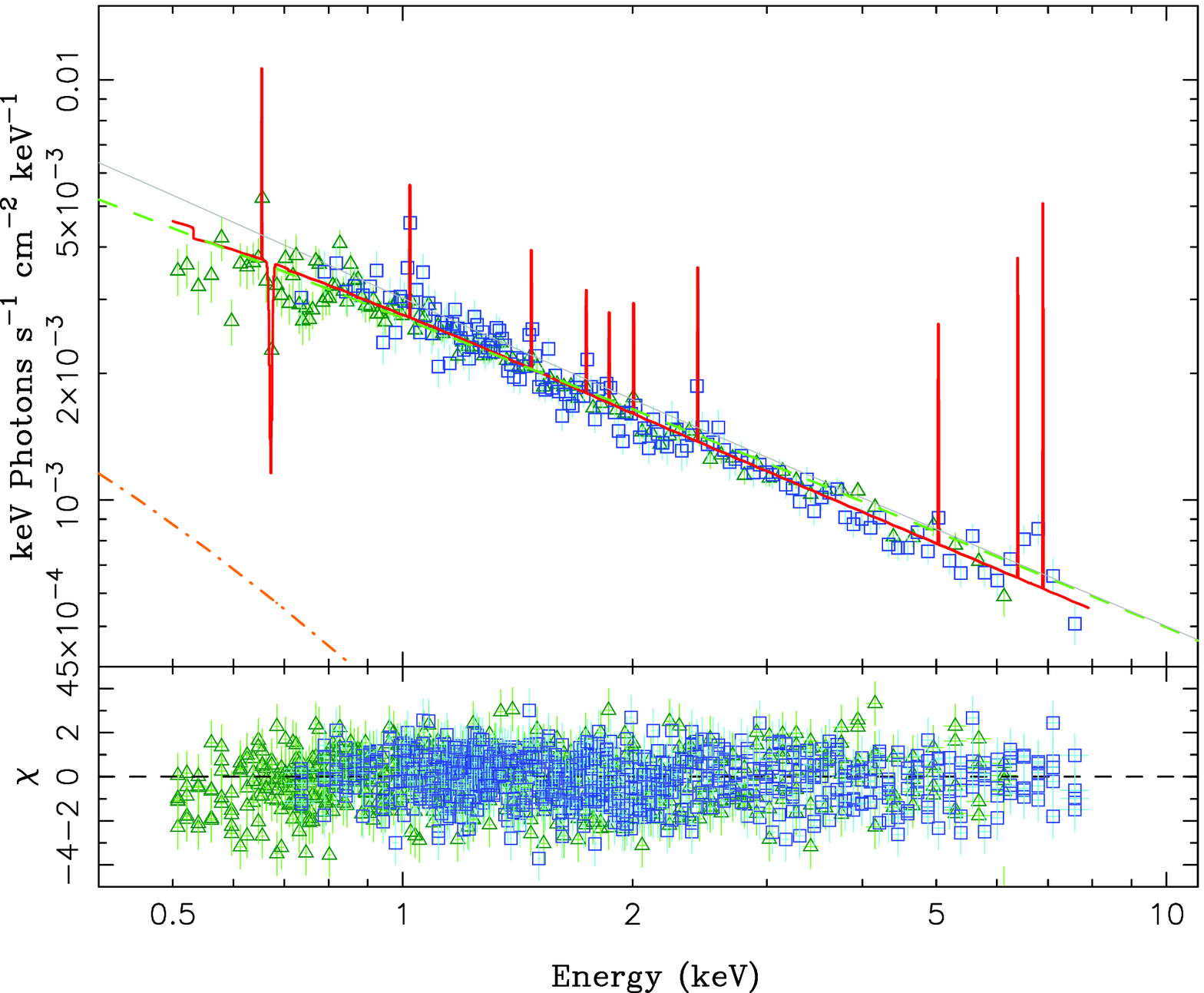}{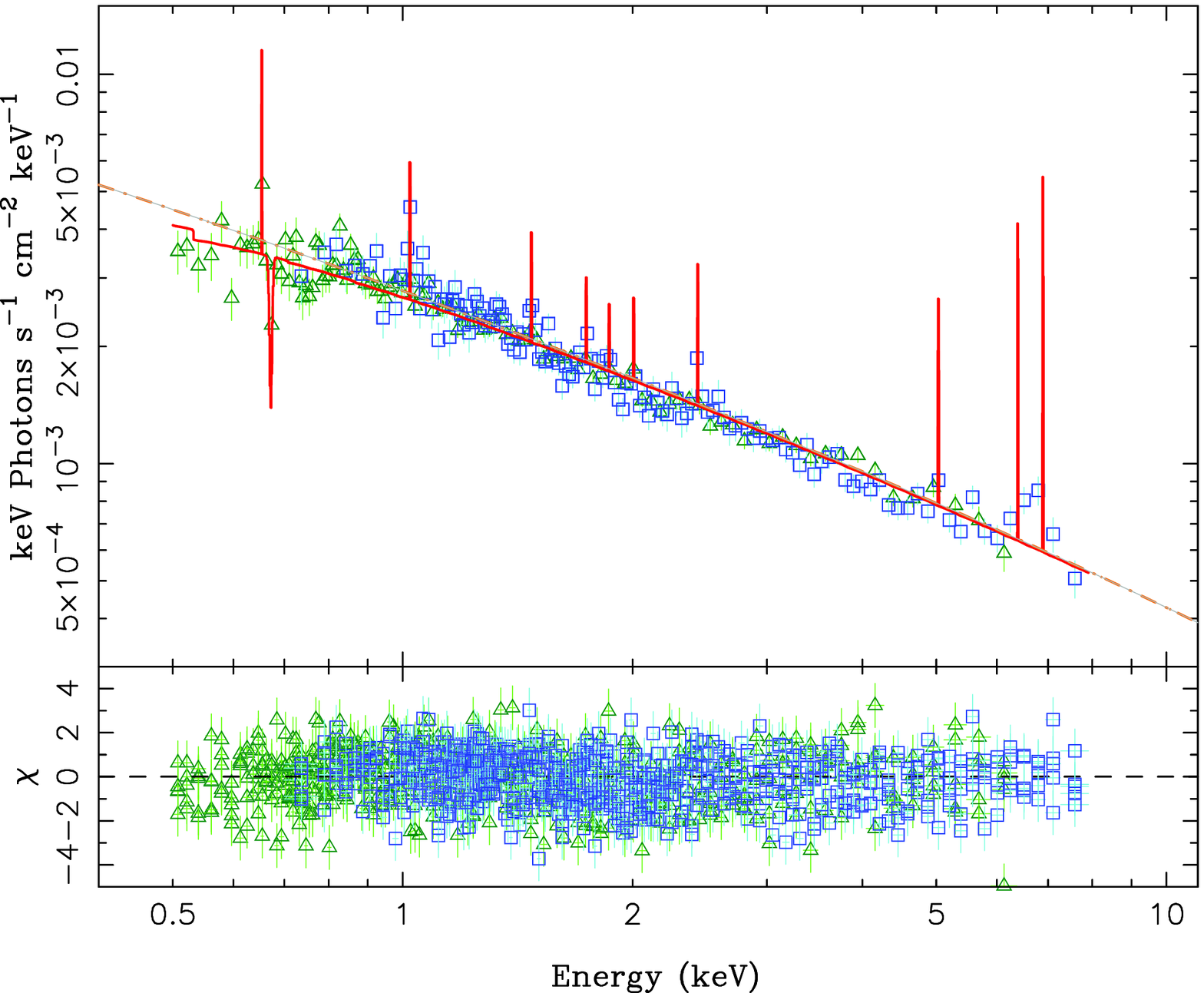}
\caption{Model fit to combined set of all broadband observations,
  closeup view of X-ray band.  Same symbol/line definitions as Fig.~\ref{febfit_xray}.  \label{mongo_x}}
\end{figure*}

\clearpage

\begin{figure*}
\epsscale{1}
\plottwo{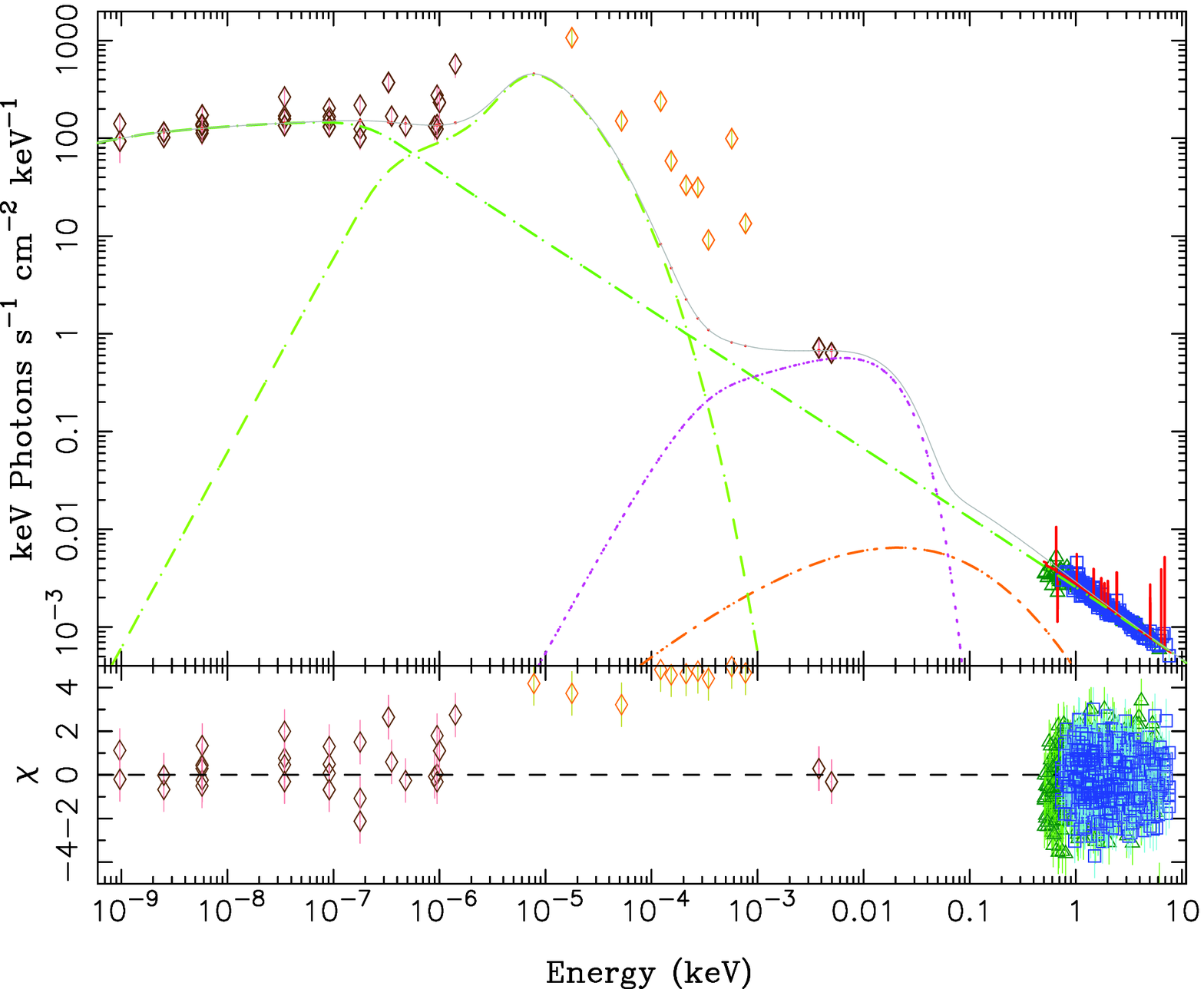}{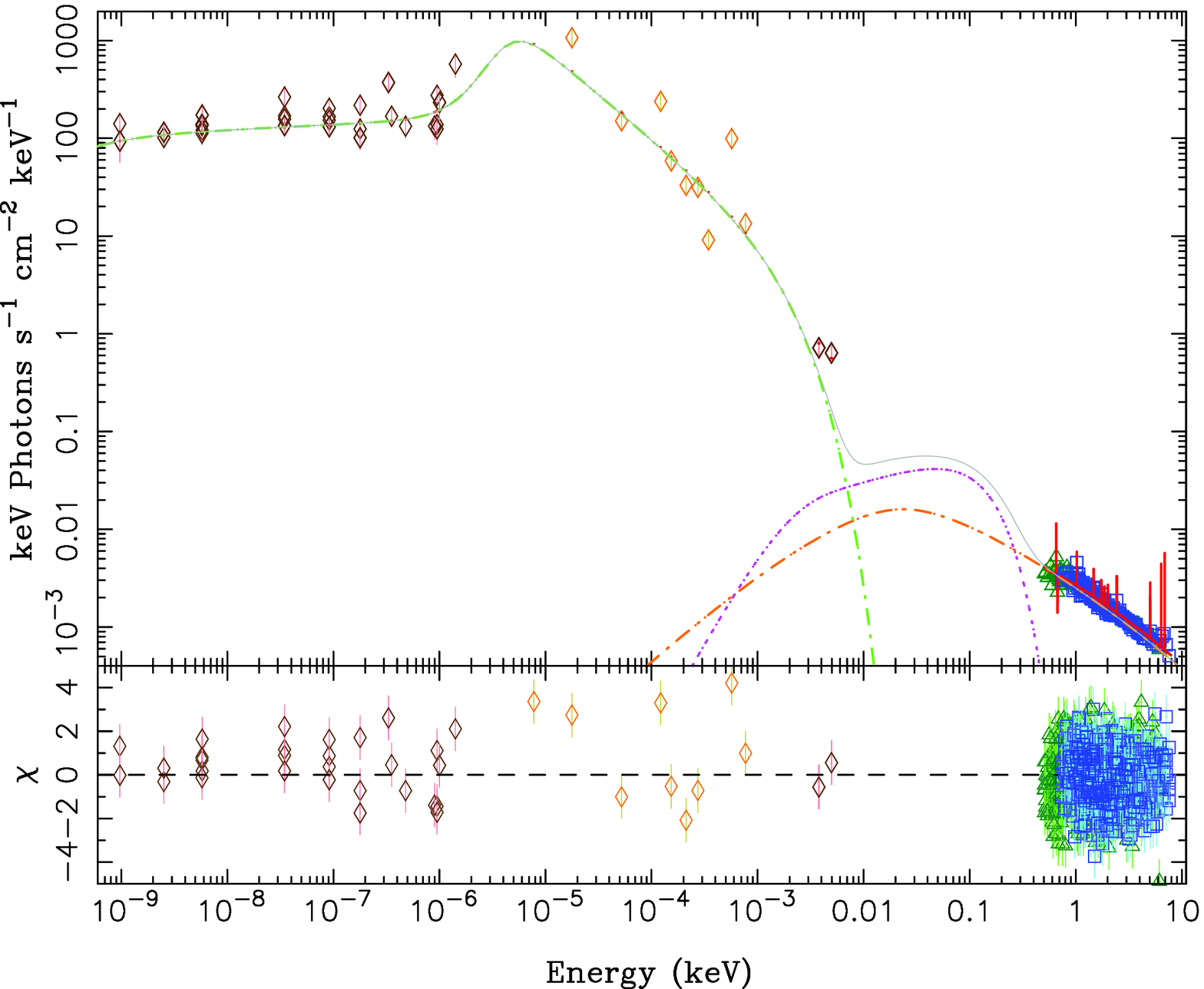}
\caption{Model fit to combined set of all broadband observations, with
  the addition of several {\em nonsimultaneous} upper limits taken
  from the literature from \textit{HST}, \textit{Spitzer}, ISO and
  MIRLIN \citep{Grossanetal2001,Gordonetal2004,Murphyetal2006}
  included as data points with 20\% systematic
  errors. Same symbol/line definitions as Fig.~\ref{febfit}.\label{mongo_ul}}
\end{figure*}

\begin{figure*}
\epsscale{1}
\plottwo{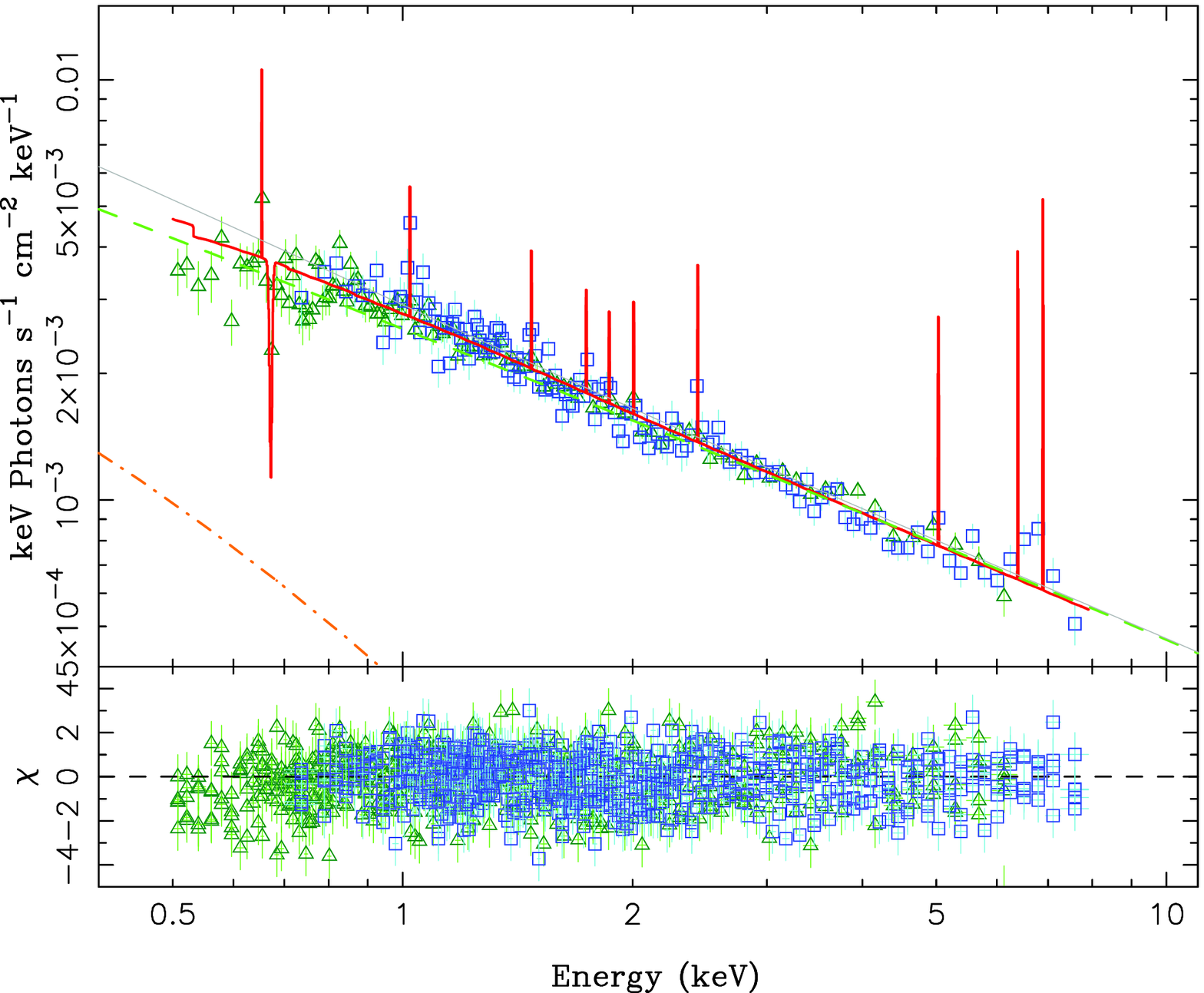}{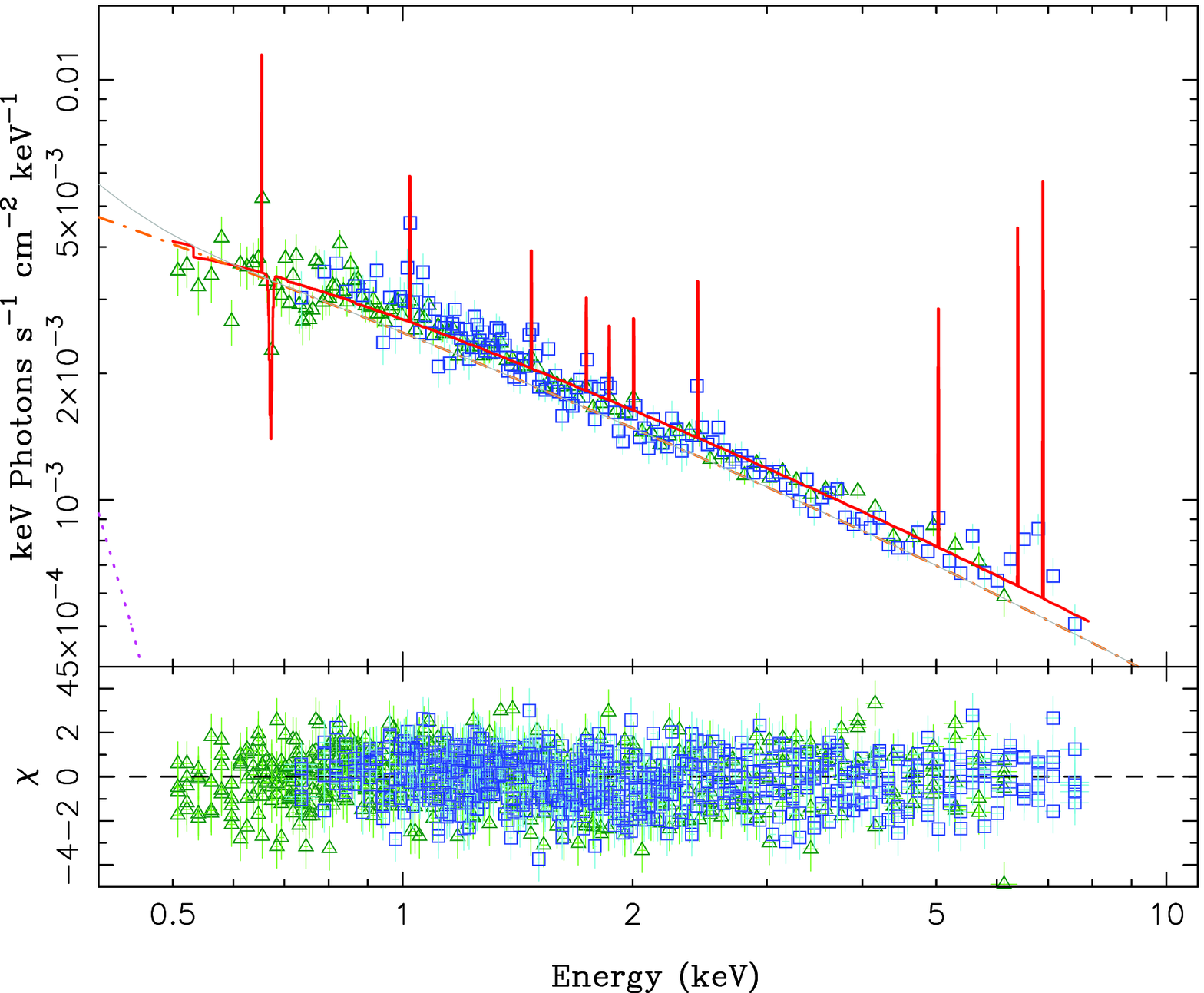}
\caption{Same as Fig.~\ref{mongo_ul}, closeup view of X-ray band. Same symbol/line definitions as Fig.~\ref{febfit_xray} \label{mongo_ul_x}}
\end{figure*}

\clearpage

\begin{figure*}
\epsscale{1}
\plotone{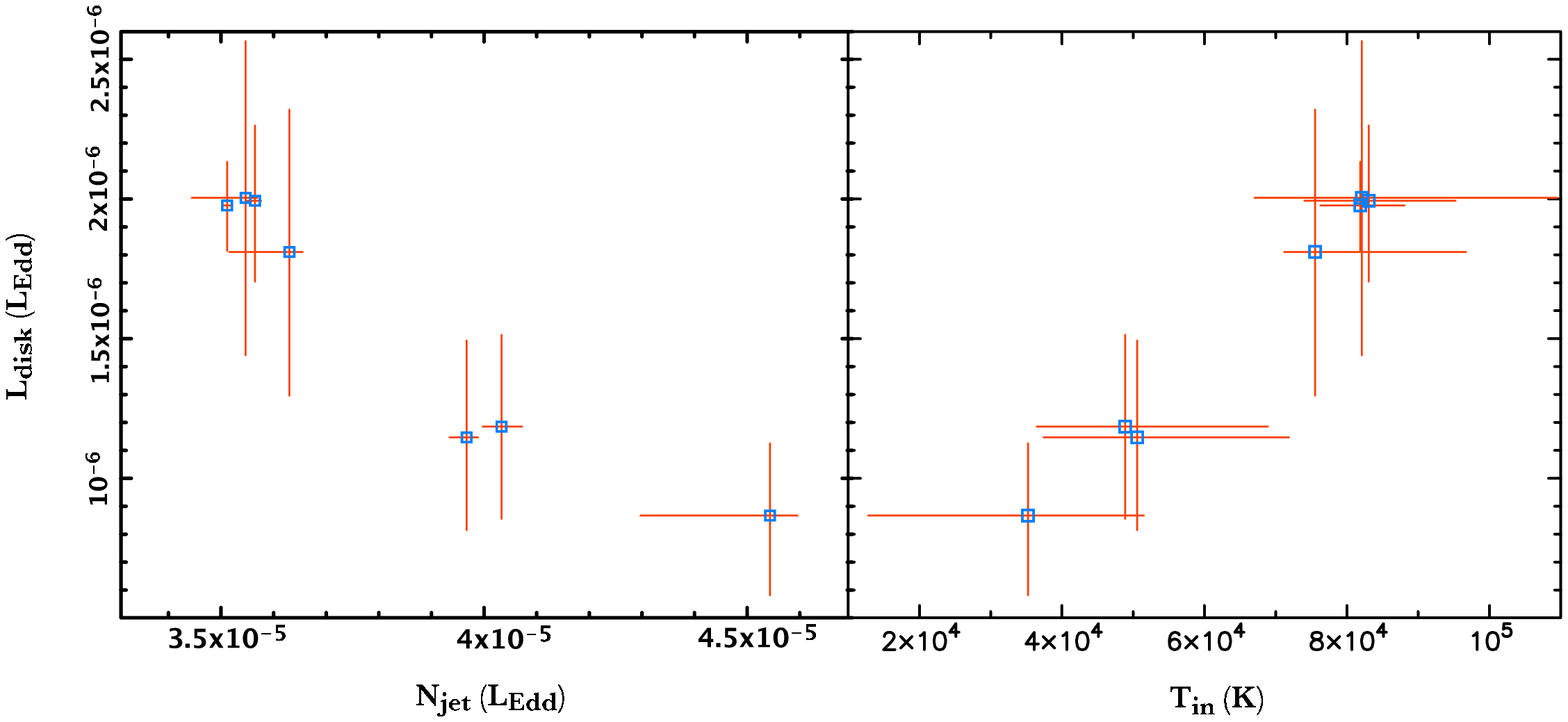}
\caption{Possible correlations between the total accretion disk
  luminosity and the jet normalization and inner disk temperature,
  based on the Maxwellian model fits to separate and combined data sets described in
  Figs.~\ref{febfit}--\ref{mongo_ul_x} and
  Tables~\ref{tab:MXSW_FINAL_a}--\ref{tab:MXSW_FINAL_b}.
\label{eddratpars} }
\end{figure*}

\begin{figure*}
\epsscale{1}
\plotone{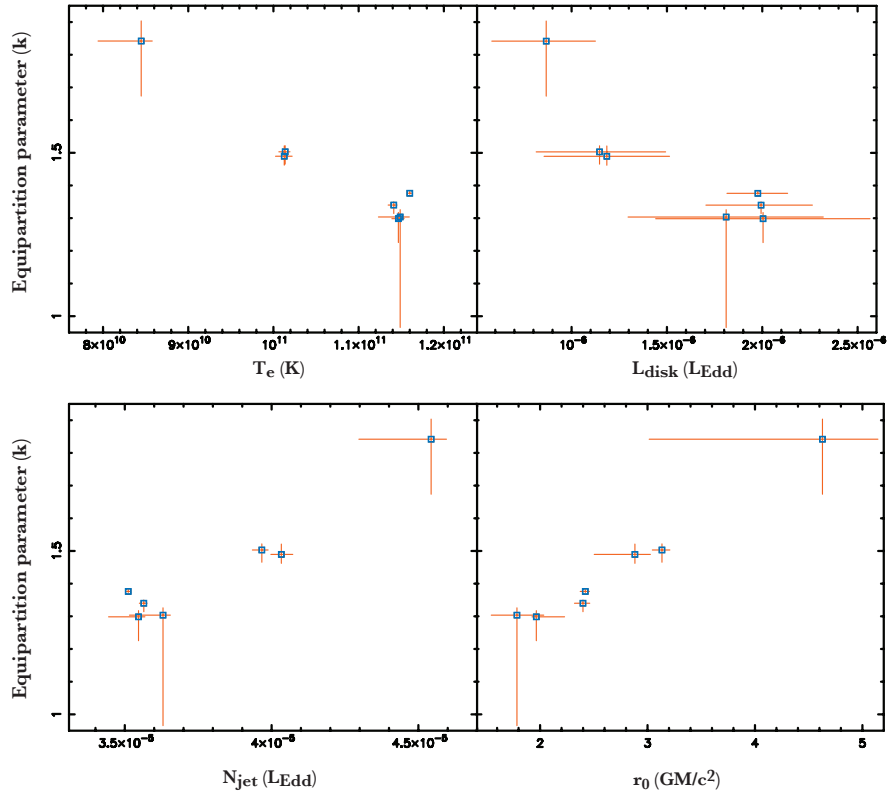}
\caption{Possible correlations between the magnetic/particle energy
  density equipartition parameter $k$ and four other fit parameters
  based on the Maxwellian fits to separate and combined data sets described in
  Figs.~\ref{febfit}--\ref{mongo_ul_x} and
  Tables~\ref{tab:MXSW_FINAL_a}--\ref{tab:MXSW_FINAL_b}.  
\label{equippars} }
\end{figure*}

\end{document}